\documentclass[aps,twocolumn]{revtex4}
\usepackage{eurosym}
\usepackage{amsfonts}
\usepackage{amsmath}
\usepackage{amssymb,epsf}
\usepackage{color}
\usepackage{graphicx}
\usepackage{epstopdf}
\usepackage{float}
\usepackage{caption}
\usepackage{subfig}

\begin{document}

\title{Structure of magnetized strange quark star in perturbative QCD}
\author{J. Sedaghat$^{1}$\footnote{%
email address: J.sedaghat@shirazu.ac.ir}, S. M. Zebarjad$^{1,2}$\footnote{
email address: zebarjad@shirazu.ac.ir}, G. H. Bordbar$^{1,3}$\footnote{%
email address: ghbordbar@shirazu.ac.ir}, and B. Eslam Panah$^{4,5,6}$\footnote{%
email address: eslampanah@umz.ac.ir}}
\affiliation{$^{1}$ Department of Physics, Shiraz University, Shiraz 71454, Iran\\
$^{2}$ Department of Physics, University of California at San Diego,\\
La Jolla, CA 92093, USA\\
$^{3}$ Department of Physics and Astronomy, University of Waterloo,\\
200 University Avenue West, Waterloo, Ontario, N2L 3G1, Canada\\
$^{4}$ Department of Theoretical Physics, Faculty of Science, University of
Mazandaran, P. O. Box 47416-95447, Babolsar, Iran\\
$^{5}$ ICRANet-Mazandaran, University of Mazandaran, P. O. Box 47416-95447,
Babolsar, Iran\\
$^{6}$ ICRANet, Piazza della Repubblica 10, I-65122 Pescara, Italy}

\begin{abstract}
We have performed the leading order perturbative calculation to obtain the
equation of state (EoS) of the strange quark matter (SQM) at zero
temperature under the magnetic field  $B=10^{18}\ G$. The SQM comprises two massless quark flavors (up and down) and one massive quark flavor(strange). Consequently, we have used the obtained EoS to calculate the maximum gravitational mass and the corresponding radius of the magnetized strange quark star (SQS). We have employed two approaches, including the regular perturbation theory (\textbf{RPT}) and the background perturbation theory (\textbf{BPT}). In \textbf{RPT} the infrared (IR) freezing effect of the coupling constant has not been accounted for, while this effect has been included in \textbf{BPT}. We have obtained the value of the maximum gravitational mass to be more than three times the solar mass.
The validity of isotropic structure calculations for SQS has also been investigated.  Our results show that the threshold magnetic field from
which an anisotropic approach begins to be significant lies in the interval $2\times 10^{18}G<B<3\times 10^{18}G$.
Furthermore, we have computed the redshift,
compactness and Buchdahl-Bondi bound of the SQS to show that this compact object cannot be a black hole.
\end{abstract}

\maketitle

\section{Introduction}

In the last stage of stellar evolution, due to the completion of the
fusion reactions in a luminous star, the gravity pressure dominates the
fusion pressure. This can lead to the birth of a highly dense matter star known as
compact star \cite%
{shapiro2008black,schmitt2010dense,glendenning2012compact,karttunen2016fundamental}%
. Having large compactness (the ratio of mass to the radius of a star) is
a feature that distinguishes compact stars from ordinary stars. Based on the range of the compactness, compact stars are classified into three groups; white dwarfs, neutron stars, and stellar black holes \cite%
{camenzind2007compact,rezzolla2018physics}. After Murray Gell-Mann \cite%
{gell2010schematic} and George Zweig \cite{zweig1964_3}, proposed quarks and
gluons as confined constituents of hadrons and also after the first experimental
evidence for this postulate in SLAC \cite{friedman1972deep}, Ivanenko,
Kurdgelaidze and then Itoh discussed a new phase
denser than nuclear matter. Such dense matter can be created by squeezing nucleons at the core of neutron
stars. This phase in which the quarks are considered as the degrees of freedom
\cite{ivanenko1965hypothesis,itoh1970hydrostatic,haensel1986strange} was
named quark matter. If the quark matter includes strange quarks, it is
called strange quark matter. Bodmer and Witten independently discussed an
interesting feature of SQM in Refs. \cite%
{bodmer1971collapsed,witten1984cosmic,farhi1984strange}. They claimed that
SQM is the true ground state of QCD which means that the minimum energy per
baryon in SQM is lower than that of the most stable nuclear matter ($%
^{56}Fe\sim 930\ MeV$). Since the strange hadrons are so unstable, to find a
stable SQM from nuclear matter, the conversion rate should be so
that a large number of quarks change their flavor to strange by weak
interactions in a short time. This situation can only be available in
the bulk size and extreme pressure, which is accessible at the core of the neutron stars \cite{farhi1986physics}. For this reason, the possibility of the existence of SQSs has been greatly discussed in literature \cite%
{alcock1986strange,chakrabarty1991equation,madsen1999physics,weber2005strange}%
. However, in Ref. \cite{holdom2018quark}, there is a discussion on a
phenomenological model including scalar and pseudoscalar nonets which results in finding a two flavor quark matter to be more stable
than SQM for baryon number $A>300$.

Nowadays, the quark matter is proposed to
exist in two new classes of compact stars: i) hybrid stars with quark
core surrounded by hadronic shell \cite%
{schertler1998influence,blaschke2001cooling,burgio2003maximum,alford2005hybrid}
and ii) pure quark stars or strange stars in which surface density exceeds
nuclear saturation density \cite%
{michel1988quark,drago2001quark,bombaci2008quark}. The behavior of  the mass-radius diagram in pure quark stars is  different from that of
neutron stars. It is originated from their different EoSs \cite%
{farhi1986physics,greiner2013nuclear,panah2019contraction,sedaghat2021compact}.%.

Until now, it has been confirmed observationally that pulsars and magnetars
have a strong magnetic field at their surface as large as $10^{12}-10^{13}\ G
$ and $10^{14}-10^{16}\ G$, respectively \cite%
{kouveliotou1998x,turolla2015magnetars}. However, theoretically, the maximum strength of the magnetic field in a magnetar is of the order of $10^{18}\ G$ \cite{lai1991cold,chanmugam1992magnetic,lai2001matter} and for self-bound
quark stars, this limit may increase up to $10^{20}\ G$ \cite%
{ferrer2010equation}. The origin of such strong magnetic fields has not been
clarified. They might be due to the fossil magnetic field of the progenitor
star \cite{chanmugam1992magnetic,sotani2015massive}. In rotating protoneutron stars in which the period of rotation is less than $3ms$, a
weak magnetic field can be intensified up to $10^{15}\ G$ \cite%
{duncan1992formation,sotani2015massive}. In such strong magnetic fields, the
Landau quantization of the transverse momenta in stellar structure is
significant \cite{landau1956zhetf,landau1977quantum,kayanikhoo2020influence}.

The magnetic field geometry in the star is still unknown. Many authors have considered a uniform magnetic field in the star \cite%
{chakrabarty1996quark,sotani2015massive,fogacca2016compact}, while others have used a density-dependent magnetic field \cite%
{dexheimer2017magnetic,sotani2017quark,kayanikhoo2020influence}. The
magnetic fields used in these references are in the range from $5\times
10^{13}\ G$ to  $10^{20}\ G$. For example, in the Ref. \cite%
{chakrabarty1996effect}, by using the MIT bag model, the maximum gravitational
mass of the SQS in the presence of a uniform magnetic field of the order of $%
10^{18}\ G$, is obtained $1.5M_{\odot }$. In the Ref. \cite%
{sotani2015massive}, the maximum mass of an isotropic massive hybrid star in
the presence of the magnetic field $B=10^{19}\ G$, has been evaluated $%
2.8M_{\odot }$. In  Ref. \cite{kayanikhoo2020influence}, the maximum mass
of the SQS under a Gaussian dependent magnetic field with the values $%
B=5\times 10^{18}\ G$ for interior and $B=5\times 10^{13}\ G$ for
the surface of the star has been calculated $1.44M_{\odot }$. Moreover, in the
Ref. \cite{sotani2017quark}, using the bag model with a density-dependent
magnetic field, the maximum mass of anisotropic hybrid star reaches to $%
2.2M_{\odot }$ for magnetic fields greater than $B=6.86\times 10^{19}\ G$ at
the core of the star.

The anisotropic effect in the calculation of the structural properties of SQS is
usually considered at strong magnetic fields. In a strong magnetic field, the rotational symmetry is broken due to the difference between the longitudinal
pressure, $P_{\lVert }$ (which is parallel to the magnetic field) and the
transverse pressure, $P_{\bot }$ (which is perpendicular to the magnetic
field) \cite{ferrer2010equation,chu2018quark}. If this difference is
negligible, the isotropic approximation can be justified; otherwise,
considering the anisotropic structure for the star is significant.

In  Ref. \cite{andreichikov2013asymptotic}, it has been shown that the running coupling constant of QCD has asymptotic
freedom in strong magnetic fields. Therefore, considering the possibility of
performing perturbative calculations due to this feature of coupling
constant in strong magnetic fields would be interesting. Thus we use
the perturbative QCD calculation to obtain the maximum gravitational mass
and the corresponding radius of SQS in the presence of the magnetic field $%
B=10^{18}\ G$ (\textcolor{red}{equipartition limit}).
In order to simplify the perturbative QCD calculations,  we consider a uniform magnetic field.
Employing QCD perturbation theory is one of the approaches to obtain the EoS
of the SQM in the absence of a magnetic field. Previously, it has been done in
Refs. \cite{fraga2005role} and \cite{kurkela2010cold}, of the order of $%
\alpha _{s}$ and $\alpha _{s}^{2}$, respectively, in the absence of a magnetic
field. In this paper, we consider the interaction between quarks and
strong magnetic field non-perturbatively and QCD interaction of quarks
perturbatively of the order of $O(\alpha _{s})$. By using the behavior of
the running $\alpha _{s}$ in the presence of a constant strong magnetic field
\cite{andreichikov2013asymptotic}, we calculate the EoS of a stable strange
quark star exposed to a uniform strong magnetic field as large as $10^{18}\ G$
at zero temperature. We then evaluate the structural properties of this star.

\textcolor{red}{In this paper, we study a pure SQS. However, in a realistic description of SQS, one should consider a hadronic layer at the star's surface. An electric dipole layer is formed in the gap between the SQM and hadronic matter. This electric dipole gap can create a strong electric field to enable the SQM to carry the hadronic layer. A discontinuity is expected between the SQM phase and the hadronic layer across the electric dipole gap \cite{weber2005strange}}. The outline of this paper is as follows. In the next section, the Landau
levels are considered as the energy states emerging in solving the Dirac
equation for a charged particle moving in a uniform magnetic field. In
section \ Ref {sec 3}, to perform perturbative calculations, the Feynman rules
for quark propagator and running coupling in the presence of uniform magnetic
field are presented. Then, the thermodynamic potential of SQM is computed
numerically up to the leading order of coupling constant. Also, the stability
conditions to obtain the EoS of the SQS in our different
approaches are explained. Two models including regular perturbation theory (\textbf{RPT}) and background perturbation theory (\textbf{BPT}) are considered. In section \ Ref {Thermodynamic
properties}, our results from perturbative calculations for the EoS and
thermodynamic properties of the SQS in \textbf{RPT} and \textbf{BPT} are
presented. Then, by numerically solving the Tolman-Oppenhaimer-Volkoff (TOV)
equations, the maximum gravitational mass and the corresponding radius of SQS in
\textbf{RPT} and \textbf{BPT} are obtained. Furthermore, the redshift,
compactness, and Buchdahl-Bondi bound of the SQS are evaluated to show that
the compact object under study cannot be a black hole. To evaluate the
validity of isotropic approximation in the calculation of the structure of SQS, the effect of
anisotropy on the structure of SQS for different magnetic fields is
investigated in section \ Ref {anisotropic structure}. We finish our paper
with some concluding remarks.

%%%%%%%%%%%%%%%%%%%%%%%%%%%%%%%%%%%%%%%%%%%%%%%%%%%%%%%%%%%%%%%%%%%%%%%%%%%%%%%%%%%%%%%%%

\section{Landau levels}

Since we are working on a QCD dense matter in a uniform strong magnetic field, we can neglect the QED interaction between quarks. In this case, there are two significant interactions, including the interaction between quarks and magnetic field and the QCD interaction between quarks. The first interaction is not perturbative, while the QCD interaction is perturbative in the high energy limit. It is natural to solve the Dirac equation for a single particle with charge $e_{f}$ and mass $m_{f}$ moving in a uniform magnetic field along the \emph{z} axis. The energy spectrum is given by
\begin{equation}
E_{j}(p_{z},B)=\sqrt{p_{z}^{2}+m_{f}^{2}+2j\lvert e_{f}B\rvert},  \label{1}
\end{equation}
where numbers, $j=0,1,2,{...}$, are referred to as Landau levels and $f$
denotes the quark flavor. It is shown that except for the lowest Landau level (LLL), the solutions are twofold degenerate for a particle with spins up and down \cite{bhattacharya2007solution}. While the orbital motion of the quarks in transverse direction, $p_{\perp }^{2}=p_{x}^{2}+p_{y}^{2}$, is quantized, the longitudinal component of the momentum, $p_{\lvert \lvert}=p_{z}$, is unaffected by the magnetic field and remains free \cite{shovkovy2013magnetic}. The splitting gap between levels is of the order of $\sqrt{2\lvert e_{f}B\rvert }$ \cite{kojo2013renormalization}. If $\mu $ is the chemical potential of the fermion, the highest Landau level is obtained by \cite{fogacca2016compact}
\begin{equation}
j_{max}=\left[ \dfrac{\mu ^{2}-m_{f}^{2}}{2e_{f}B}\right] ,  \label{2}
\end{equation}%
where $[x]$ is the integer part of $x$.

\section{Fermion propagator and QCD running coupling in uniform magnetic field}

\label{sec 3}

To calculate the perturbative contribution of the Feynman diagrams, we need to know fermion and gluon propagators in the presence of a strong magnetic field. It is more convenient to work in Euclidean space at the finite chemical potential. So the Euclidean metric $(g_{\mu\nu} = \delta_{\mu\nu})$ must be used. This makes no difference between the covariant and contravariant components of the
vectors. In Euclidean space, the propagator of a quark with charge $e_{f}$ moving in a uniform magnetic field is given in terms of generalized Laguerre polynomials labeling the $j$th Landau levels \cite{miransky2015quantum}.
\begin{equation}
S(p)=-i\sum_{j=0}^{\infty }\left( -1\right) ^{j}e^{\frac{-p_{\perp }^{2}}{{%
			\lvert e_{f}B\rvert }}}\dfrac{D_{j}\left( \lvert e_{f}B\rvert ,p\right) }{%
	p_{0}^{2}+\left( p_{3}^{2}+m_{f}^{2}+2j\lvert e_{f}B\rvert \right) },
\label{propagator of quark}
\end{equation}%
where,
\begin{align}
D_{j}\left( \lvert e_{f}B\rvert ,p\right) =& (m_{f}-\gamma _{0}p_{0}-\gamma
_{3}p_{3})  \notag \\
&  \notag \\
& \times \left[ (1+\mathcal{K})L_{j}\left( \mathcal{X}\right) -(1-\mathcal{K}%
)L_{j-1}\left( \mathcal{X}\right) \right]  \notag \\
&  \notag \\
& +4\left( \gamma _{1}p_{1}+\gamma _{2}p_{2}\right) L_{j-1}^{(1)}\left(
\dfrac{2p_{\bot }^{2}}{e_{f}B}\right) .  \label{lauguerre}
\end{align}%
where $\gamma _{i}$s are Dirac gamma matrices in Euclidean space. Also, $%
\mathcal{K}=$ $i\gamma _{1}\gamma _{2}$sign$(e_{f}B)$ and $\mathcal{X=}%
\dfrac{2p_{\bot }^{2}}{e_{f}B}$. At finite chemical potential, the zero
components of the fermionic momenta are shifted by $p_{0}\longrightarrow
p_{0}+i\mu $ \cite{kurkela2010cold}. Since gluon has no
interaction with magnetic field, its propagator has the same regular form as in
absence of magnetic field. Thus in Feynman gauge, the gluon propagator is
\begin{equation}
D_{\mu \nu }(k)=\dfrac{\delta_{\mu \nu }}{k^{2}}.
\end{equation}

The running coupling of QCD at one-loop level (one gluon exchange
approximation) in the presence of uniform magnetic field has been given in  \cite{andreichikov2013asymptotic}
\begin{equation}
\alpha (Q,B)=\frac{\alpha (\Lambda _{0})}{1+\alpha (\Lambda _{0})\left[
\dfrac{11N_{c}}{12\pi }\ln \left( \dfrac{Q^{2}+M_{B}^{2}}{\Lambda _{0}^{2}}%
\right) +F(Q,B)\right] },  \label{5}
\end{equation}%
where
\begin{eqnarray}
F(Q,B) &=&\dfrac{1}{Q^2}\sum_{f}\dfrac{\lvert e_{f}B\lvert }{\pi }\exp \left( \dfrac{%
-q_{\bot }^{2}}{2e_{f}B}\right) T\left( \dfrac{Q^{2}}{4m_{f}^{2}}\right) , \\
&&  \notag \\
\alpha (\Lambda _{0}) &=&\dfrac{12\pi }{11N_{c}\ln \left( \dfrac{\Lambda
_{0}^{2}+M_{B}^{2}}{\Lambda _{V}^{2}}\right) },  \label{B}
\end{eqnarray}%
and
\begin{equation}
T(z)=-\dfrac{\ln (\sqrt{1+z}+\sqrt{z})}{\sqrt{z(z+1)}}.
\end{equation}

With an accuracy better than $10\%$, $T(z)$ becomes \cite%
{vysotsky2010atomic,andreichikov2013asymptotic}
\begin{equation}
T(z)=\dfrac{2z}{3+2z}.
\end{equation}

In Eqs. (\ref{5}) and (\ref{B}), $N_{c}$ is the number of colors and $M_{B}$ is
originated from background perturbation theory (\textbf{BPT}) for explaining infrared (IR)
freezing effect \cite{simonov2011asymptotic,deur2016qcd}. According to this effect which is represented by some experiments such as colliding polarized electrons to protons \cite{deur2008determination} or hadronic decays of the $\tau $ lepton \cite{ackerstaff1999measurement}, the running coupling has
slowly varying behavior in the IR region of energy. IR freezing
effect can be explained in such a way that the running behavior of the
coupling constant is originated from particle-antiparticle loop corrections.
Confinement implies that quarks and anti-quarks cannot have a wavelength more than the size of the hadron. Thus the loop corrections for the coupling constant are suppressed in
IR scale and $\alpha _{s}$ is expected to lose its energy dependence \cite%
{brodsky2008maximum,deur2016qcd}. We have chosen two approaches with and
without considering the IR freezing effect in running coupling to check the impact of this QCD effect on the star structure. These two approaches are as
follows: i) without considering IR freezing effect in which $M_{B}=0$ (%
\textbf{RPT}) and ii) with considering IR freezing effect in which $M_{B}$ $%
\cong 1\ GeV$ interpreted as the ground-state mass of two gluons connected
by the fundamental string, with string tension $\sigma =0.18\ GeV^{2}$ (%
\textbf{BPT}) \cite{andreichikov2013asymptotic,ferrer2015quark}. $\Lambda
_{V}$ is the QCD scale parameter which is considered $0.38\ GeV$ and $0.48\
GeV$ in \textbf{RPT} and \textbf{BPT}, respectively \cite{badalian2019radial}%
. We have used $\Lambda _{0}=2\ GeV$ for both approaches and considered the
strange quark as a massive particle while the masses of up and down quarks
are negligible. The running mass of strange quark up to one-loop correction in the absence of a magnetic field is given by \cite{vermaseren19974}
\begin{equation}
m_{s}(Q)=m_{s}(2GeV)\left[ \dfrac{\alpha _{s}(Q)}{\alpha _{s}(2GeV)}\right]
^{\dfrac{\gamma _{0}}{\beta _{0}}},  \label{9}
\end{equation}%
where $m_{s}(2\ GeV)\cong 0.1\ GeV$ \cite{tanabashi2018review}, and $\beta
_{0}$ and $\gamma _{0}$ are beta function and anomalous dimension,
respectively in one-loop order which are given as
\begin{eqnarray}
\beta _{0} &=&\dfrac{1}{3}\left( 11N_c-2N_{f}\right) ,  \notag \\
&&  \notag \\
\gamma _{0} &=&3C_{F},
\end{eqnarray}%
where $N_{f}$ is the number of flavors and $C_{F}$ is given as
\begin{eqnarray}
C_{F}=\dfrac{N_{c}^{2}-1}{2N_{c}}.
\end{eqnarray}

\begin{figure}[tbp]
\center{\includegraphics[width=8.5cm]
		{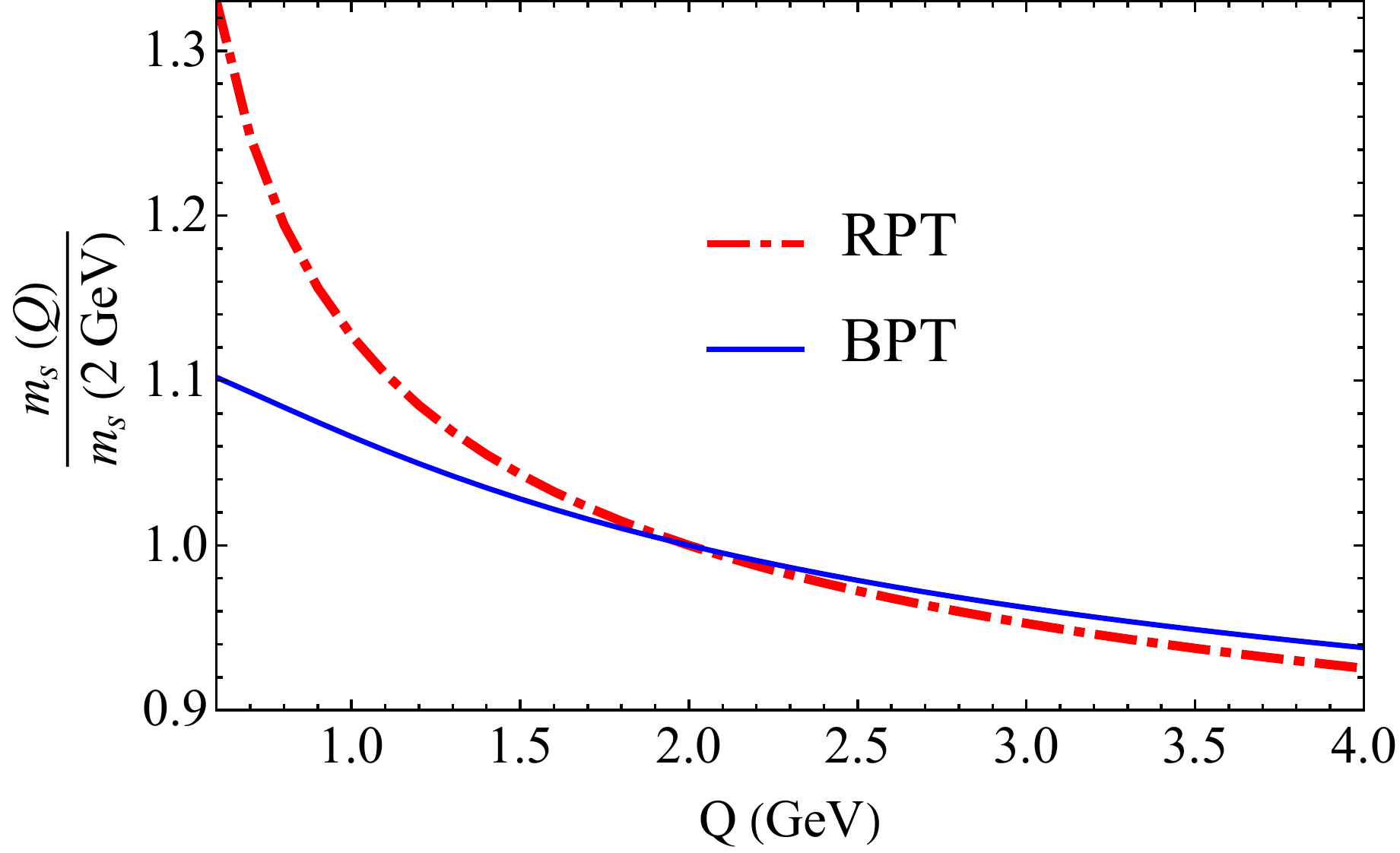}}
\caption{{\protect\small {\ Running mass of strange quark versus energy
scale ($Q$) in presence of magnetic field $B=10^{18}\ G$ in \textbf{RPT}
(full curve) and \textbf{BPT} (dashed curve) approaches.}}}
\label{Run-mass}
\end{figure}

By inserting Eq. (\ref{5}) in Eq. (\ref{9}), and assuming the same initial
conditions, we can obtain the running mass of strange quark in both \textbf{RPT} and \textbf{BPT} which is shown in Fig. \ref{Run-mass}. As we can see from Fig. \ref{Run-mass}, the running mass of the strange quark in \textbf{BPT} is lower than that of \textbf{RPT} below $Q=2\ GeV$. In energies more than this value, the running mass in \textbf{RPT} is lower than that of \textbf{BPT}.

\begin{figure}[tbp]
\center{\includegraphics[width=8.5cm]
		{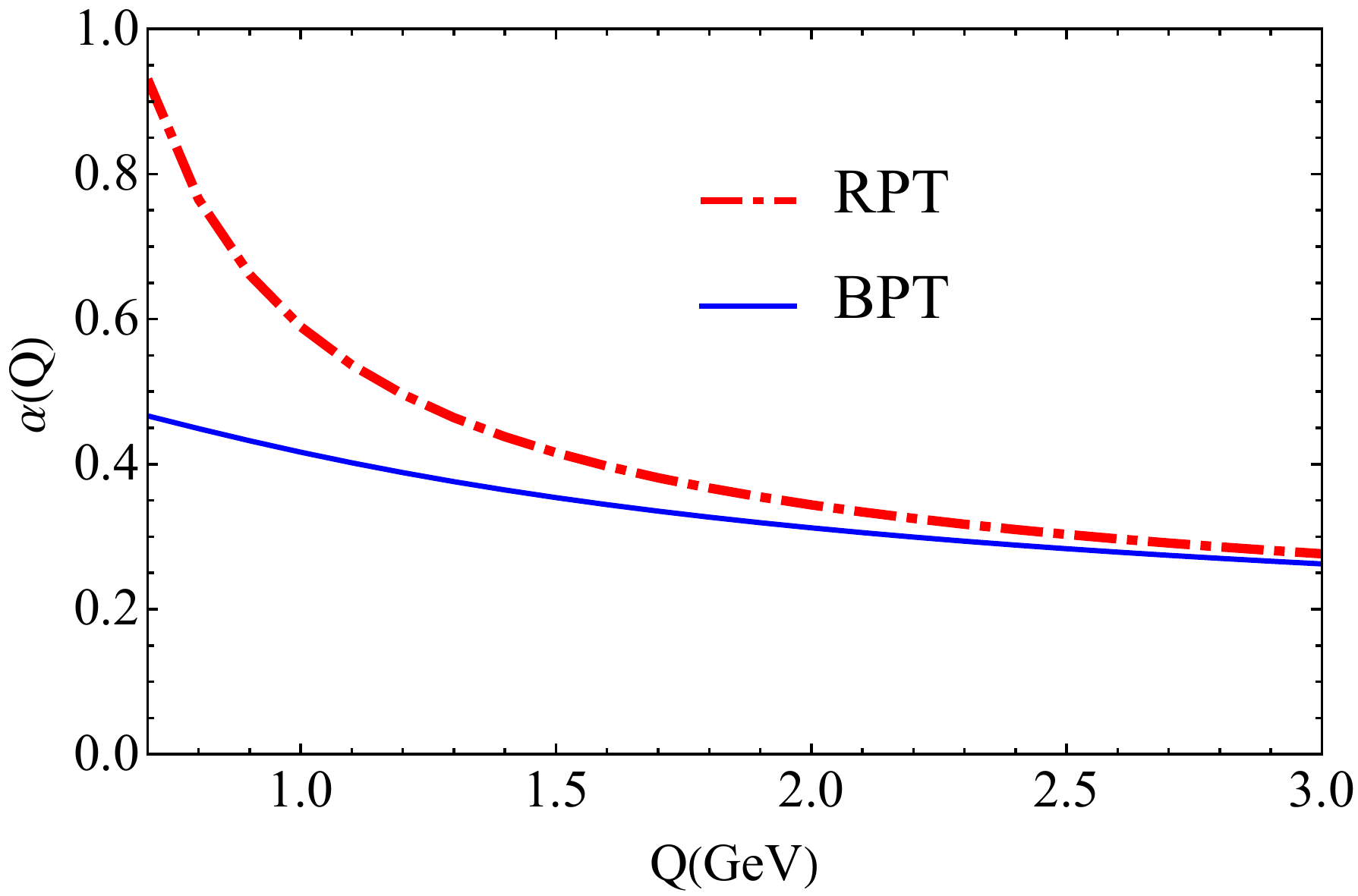}}
\caption{{\protect\small {The comparison of QCD running coupling constant in
presence of magnetic field $B=10^{18}\ G$ in \textbf{RPT} (dashed curve) and
\textbf{\ BPT} (dotted curve) approaches.}}}
\label{comparison of couplings}
\end{figure}

Fig. \ref{comparison of couplings} shows the comparison of the
running coupling constant in the presence of $B=10^{18}$ G in both\textbf{\ RPT} and \textbf{\ BPT} approaches. From this figure, we see that in all energy scales, the coupling constant in \textbf{BPT} is lower than that of \textbf{RPT}. Furthermore, the running coupling in \textbf{BPT} approach changes smoothly due to IR freezing effect.

\section{Thermodynamic potential up to the leading order}

\label{sec 4}

In grand canonical ensemble, the grand potential $\Omega $ at zero
temperature is defined by
\begin{equation}\label{eos}
\dfrac{\Omega }{V}=\epsilon -{\mu }_fn_f-{\mu }_en_e,
\end{equation}%
where $V$ and $\epsilon$ are the volume and energy density of the system, respectively.  $\mu_f$ and $n_f$ are quark's chemical potential and  density with flavor $f$. $\mu_e$ and $n_e$ are the electron chemical potential and electron number density, respectively. The quark number density, electron number density,
longitudinal pressure and transverse pressure can be evaluated from the grand potential by
\begin{eqnarray}
n_{i,e} &=&-\dfrac{1}{V}\dfrac{\partial \Omega }{\partial \mu_{i,e} },  \label{density}
\\
&&  \notag \\
P_{\lVert } &=&-\mathfrak{B}-\dfrac{\Omega }{V}+P_{CSC}-\dfrac{B^{2}}{8\pi },
\label{p} \\
&&  \notag \\
P_{\bot } &=&-\mathfrak{B}-\dfrac{\Omega }{V}+P_{CSC}-M.B+\dfrac{B^{2}}{8\pi
},  \label{p transvers}
\end{eqnarray}%
where $\mathfrak{B}$ is a free parameter interpreted as non-perturbative
effects not included in perturbative expansion \cite{kurkela2010cold}. To
describe a more realistic quark matter, we have considered the color
superconductivity pressure, $P_{CSC}$. This pressure is due to an appearing
gap in the energy of the Cooper pairs in the Color-Flavor-Locked phase
which is given as $P_{CSC}=\dfrac{\Delta ^{2}(\mu _{u}+\mu _{d}+\mu _{s})^{2}%
}{3\pi ^{2}}$, \cite{alford1999color} where  $\Delta $ is the gap parameter. The gap parameter depends on the magnetic field. Therefore,  we consider different values for $\Delta$ ranging from $0$ to $100MeV$ in appendix B. This appendix shows that the maximum gravitational mass of SQS increases by increasing $\Delta$. In the following, we only obtain our results for $\Delta = 100MeV$.
In Eq. (\ref%
{p transvers}), $M$ is the magnetization of the system, which is calculated by
$-\dfrac{\partial \Omega }{\partial B}$. To obtain the grand potential, we
use the fact that $\Omega $ for each quark flavor can be divided into two
perturbative and non-perturbative parts
\begin{equation}\label{omegafreeper}
\Omega =\Omega _{free}+\Omega _{2L},
\end{equation}%
where the non-perturbative part, $\Omega _{free}$, is the thermodynamic the potential of a free fermion moving in a uniform magnetic field which is given by
\begin{align}\label{free}
\frac{\Omega _{free}}{V}= & -\,\sum_{f,j=0}^{\infty }
\frac{|e_{f}B|}{2\pi ^{2}}\bigg\{3(2-\delta _{j0})  \notag \\
& \times \int_{0}^{(k_{z}^{f})_{F}}dk_{z}\frac{k_{z}{}^{2}}{\sqrt{%
m_{f}{}^{2}+k_{z}{}^{2}+2j|e_{f}B|}}\bigg\} \notag\\
&  -\,\sum_{j=0}^{\infty }\frac{|eB|}{2\pi ^{2}}
\bigg\{(2-\delta _{j0})  \notag \\
& \times \int_{0}^{(k_{z}^{e})_{F}}dk_{z}\frac{k_{z}{}^{2}}{\sqrt{%
		m_{e}{}^{2}+k_{z}{}^{2}+2j|eB|}}\bigg\}
\end{align}

\begin{figure}[]
	\center{\includegraphics[width=2.5cm]
		{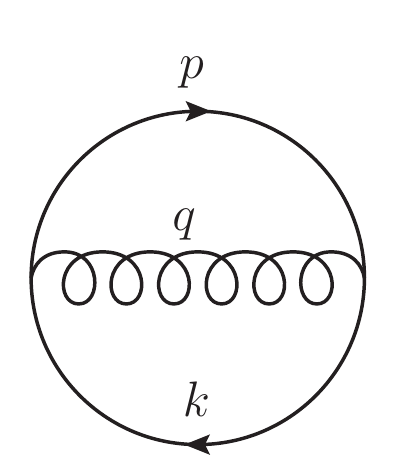}}
	\caption{\label{fig:one} \small{Two-loop diagram contributing to  perturbative part of  the grand potential at the leading order. }}
	\label{2loop}
\end{figure}

In above equation, $\left( k_{z}^{f}\right) _{F}$ denotes the quark fermi
momentum with flavor $f$ which is $\sqrt{\mu _{f}^{2}-m_{f}^{2}-2j|e_{f}B|}$. $\left( k_{z}^{e}\right) _{F}$ denotes the electron fermi
momentum with which is $\sqrt{\mu _{e}^{2}-m_{e}^{2}-2j|eB|}$. The
perturbative part of the grand potential, $\Omega _{2L}$, is the QCD
contribution at leading order for a system of quarks and gluons under strong
magnetic field which is depicted by a two-loop diagram in Fig. \ref{2loop}.
Appendix \ref{appendixA} shows the procedure to calculate the two-loop diagram at finite chemical potential.
The masses of up and down quarks are
negligible. For the mass of strange quark, we use the running mass from
Eq. (\ref{9}). The result of  two-loop diagram (Eq. (\ref{omegatotal2})), depends on the renormalization scale $%
Q $ that appears in mass and coupling constant through Eqs. (\ref{5}) and (\ref%
{9}). In the absence of strong magnetic field, phenomenological models
propose $Q=2\pi \sqrt{T^{2}+\frac{\mu ^{2}}{\pi ^{2}}}$ for a massless quark
with chemical potential $\mu $ at temperature $T$. At zero temperature, for
a system composed of $N_{f}$ flavors, $Q=2\left( {\sum_{f}\mu _{f}/3}\right)
\equiv 2\overline{\mu }$ which can have a variation by a factor of 2 with
respect to its central value \cite%
{schneider2003qcd,bandyopadhyay2019pressure,karmakar2019anisotropic}.

\begin{figure}[tbp]
	\center{\includegraphics[width=8.5cm]
		{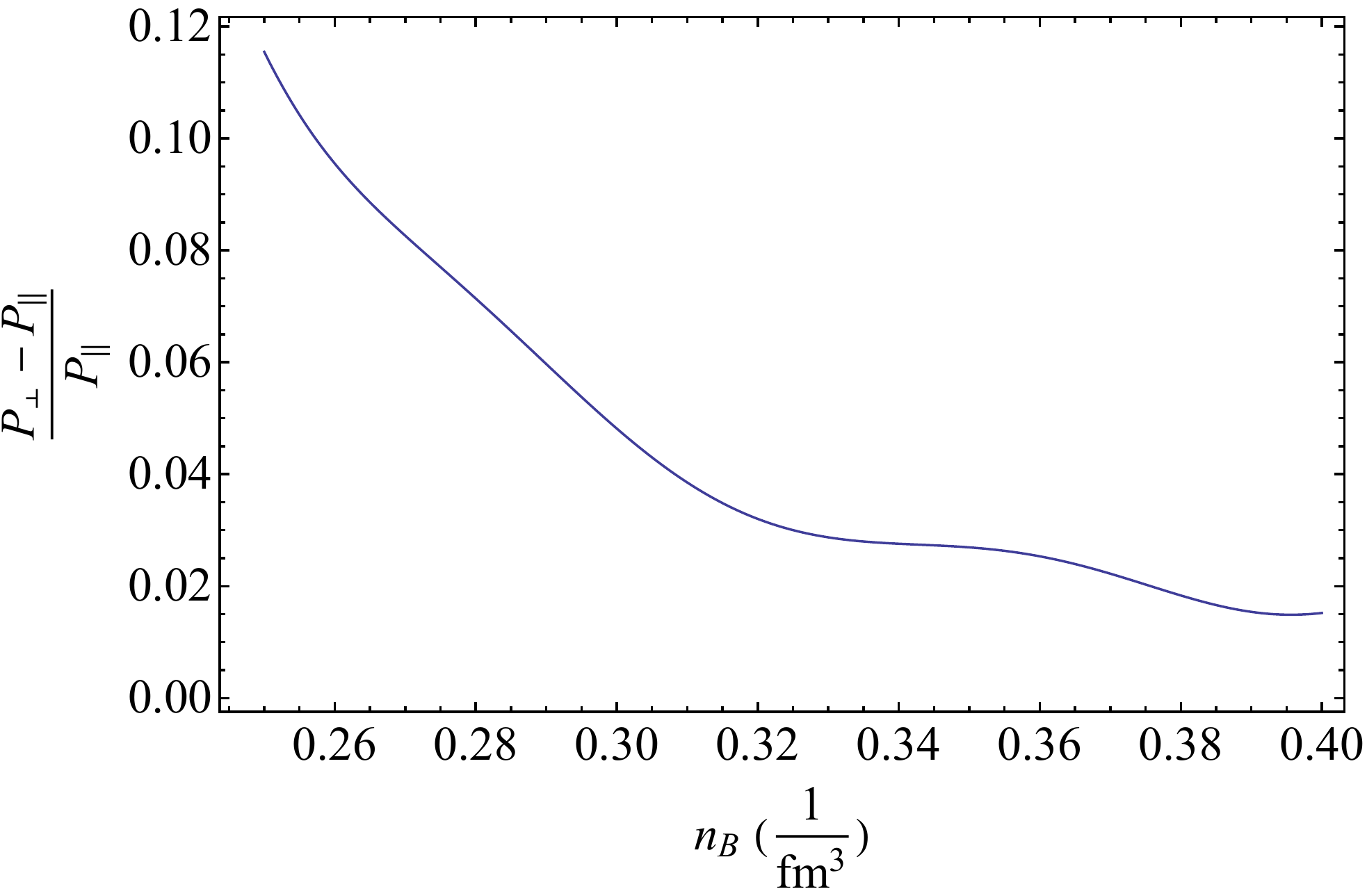}}
	\caption{{\protect\small {The ratio of $\dfrac{P_{\bot }-P_{\lVert }}{
					P_{\lVert }}$ versus baryon number density. }}}
	\label{comparing of pressures}
\end{figure}

Fig. \ref{comparing of pressures} shows the ratio of $\dfrac{P_{\bot
	}-P_{\lVert }}{P_{\lVert }}$ versus baryon density, $n_{B}$, which indicates
that the difference between longitudinal and transverse pressures decreases
by increasing baryon number density. In section \ref{anisotropic structure}, we show that the maximum gravitational masses due to these pressures ($P_{\lVert } , P_{\bot
}$) result in the same values approximately. Therefore our assumption for having an isotropic
structure for the system under study is justified.

\begin{figure}[tbp]
	\center{\includegraphics[width=8.5cm]
		{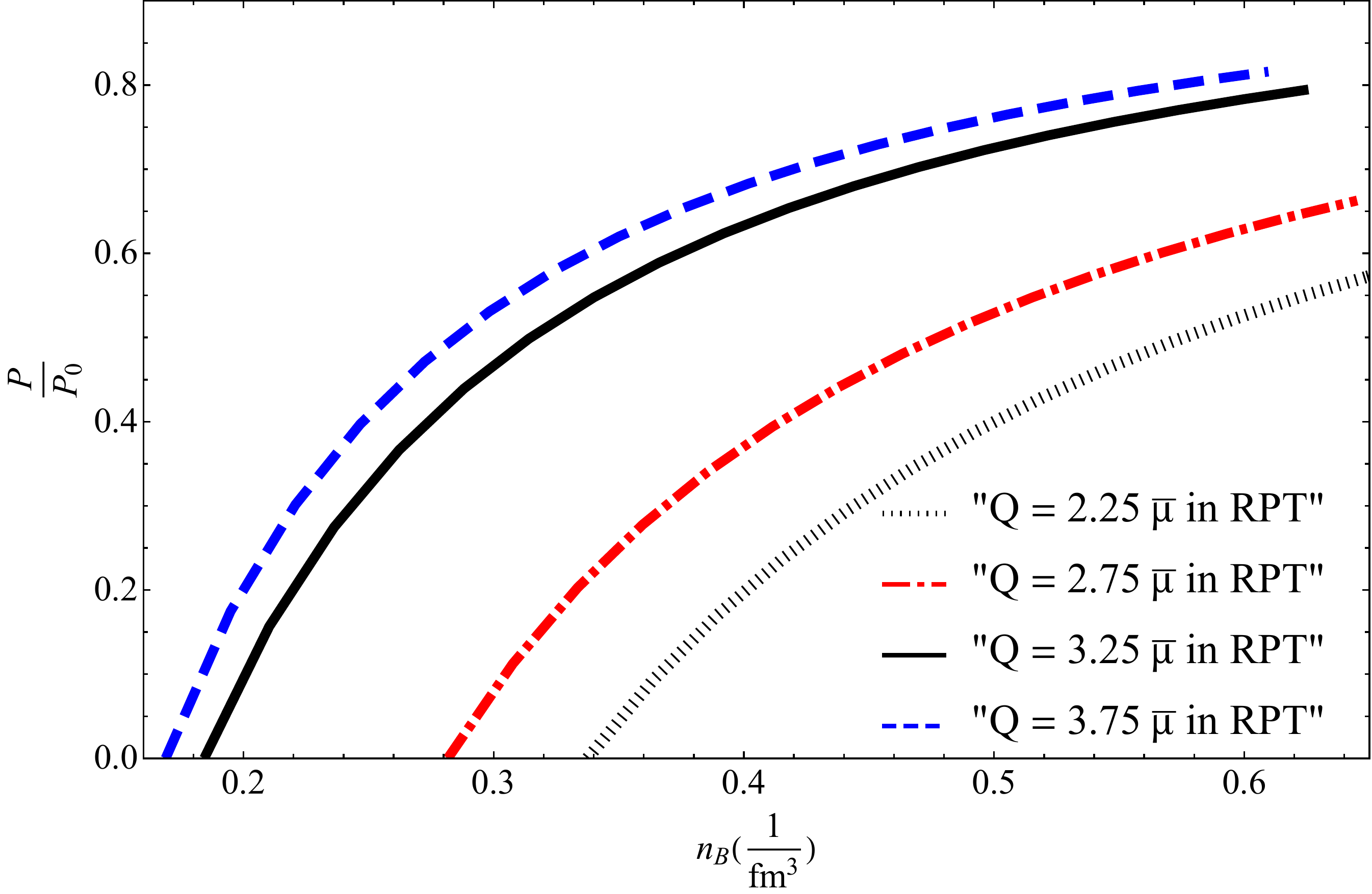}}
	\caption{{\protect\small {The ratio of calculated pressure up to leading
				order in $\protect\alpha _{s}$ to the pressure of the free quarks versus
				renormalization scale $Q$ for three choices of renormalization scale in
				\textbf{RPT} approach}.}}
\label{validity}
\end{figure}

Fig. \ref{validity} shows the validity of our perturbative
calculations. In this figure, the ratio of calculated pressure up to the leading
order in $\alpha _{s}$ to the pressure of free quarks is presented. As the
figure shows, this ratio tends to unity by increasing the renormalization
scale $Q$ for three choices of renormalization scale in \textbf{RPT}
approach. Such an asymptotic behavior is necessary for any perturbative method, which shows that the perturbative part is lower than the non-perturbative part. Also, the non-perturbative contribution decreases by increasing the energy due to the asymptotic freedom of the coupling constant. Furthermore, we have imposed the following constraints in our perturbative calculations:

i) To establish the causality, the ratio of the speed of sound to the
speed of light in the vacuum should be lower than unity.

ii) To establish dynamical stability, the adiabatic index should be lower
than $\dfrac{4}{3}$ \cite{panah2019contraction}.

iii) The relation between pressure and energy density is in such a way that $%
\epsilon + P \geq 0 $ and $\epsilon \geq \mid P \mid$ \cite{roupas2021qcd}.

iv) To describe a realistic SQS, the minimum value of the baryon number
density should be higher than the nuclear saturation density $(n_B=0.16fm^{-3})$.


\section{Equation of state and stability conditions}\label{sec 5}
In this section, we use the numerical results from the grand
potential presented in the previous section to derive the EoS of a stable
SQS. We derive the relationship
between the total energy density and pressure (EoS) to evaluate the star's maximum mass and
corresponding radius. Therefore, we can derive the EoS using Eqs. (\ref
{density}-\ref{free}) and (\ref{omegatotal2}) from the following relation
\begin{equation}
\varepsilon =-P_{\lVert }+\mu _{u}n_{u}+\mu _{d}n_{d}+\mu _{s}n_{s}.
\label{22}
\end{equation}%
It should be noted that
in deriving the EoS, we have considered beta equilibrium and charge neutrality conditions \cite{blaschke2001physics}. For beta equilibrium, we have the following weak processes.
\begin{equation*}
d\longrightarrow u+e+\bar{\nu}_{e},~~~\&~~~u+e\longrightarrow d+\nu _{e},
\end{equation*}%
\begin{equation*}
s\longrightarrow u+e+\bar{\nu}_{e},~~~\&~~~u+e\longrightarrow s+\nu _{e},
\end{equation*}%
\begin{equation}
s+u\longleftrightarrow d+u,
\end{equation}%
which lead to the following conditions
\begin{equation}
\mu _{s}=\mu _{d}\equiv \mu ,~~~\&~~~\mu _{u}=\mu -\mu _{e}.  \label{24}
\end{equation}

The mean free path of neutrinos is considerably larger than the size of the
star, and therefore their chemical potential can be neglected in the
calculations. For charge neutrality, the following condition must be imposed.
\begin{equation}
\frac{2}{3}n_{u}-\frac{1}{3}n_{d}-\frac{1}{3}n_{s}-n_{e}=0,  \label{25}
\end{equation}%
where $n_{e}$, the electron number density, is equal to
\begin{equation}
n_{e}=\dfrac{eB}{(2\pi )^{2}}\mu _{e}^{2}.  \label{26}
\end{equation}

Furthermore, for the existence of SQS, another condition must be
implemented, which implies that the {minimum} energy per baryon should be lower than that of the most stable nuclei $\left( ^{56}Fe\right) $ \cite
{weber2005strange}
\begin{equation}
\dfrac{\epsilon }{n_{B}}\leq 0.93GeV,  \label{energy constraint}
\end{equation}%
where $n_{B}$, the baryon number density, is obtained by quark number densities as
\begin{equation}
n_{B}=\dfrac{n_{u}+n_{d}+n_{s}}{3}.  \label{nB}
\end{equation}%
{}

\section{Thermodynamic properties of strange quark matter in strong magnetic
field}

\label{Thermodynamic properties}

In this section, we investigate our results for thermodynamic properties
such as baryon density, speed of sound, adiabatic index, and also the EoS of SQM in \textbf{RPT} and \textbf{BPT}, respectively.

\subsection{Results in \textbf{RPT}}

\begin{figure}[h]
\center{\includegraphics[width=8.5cm]
		{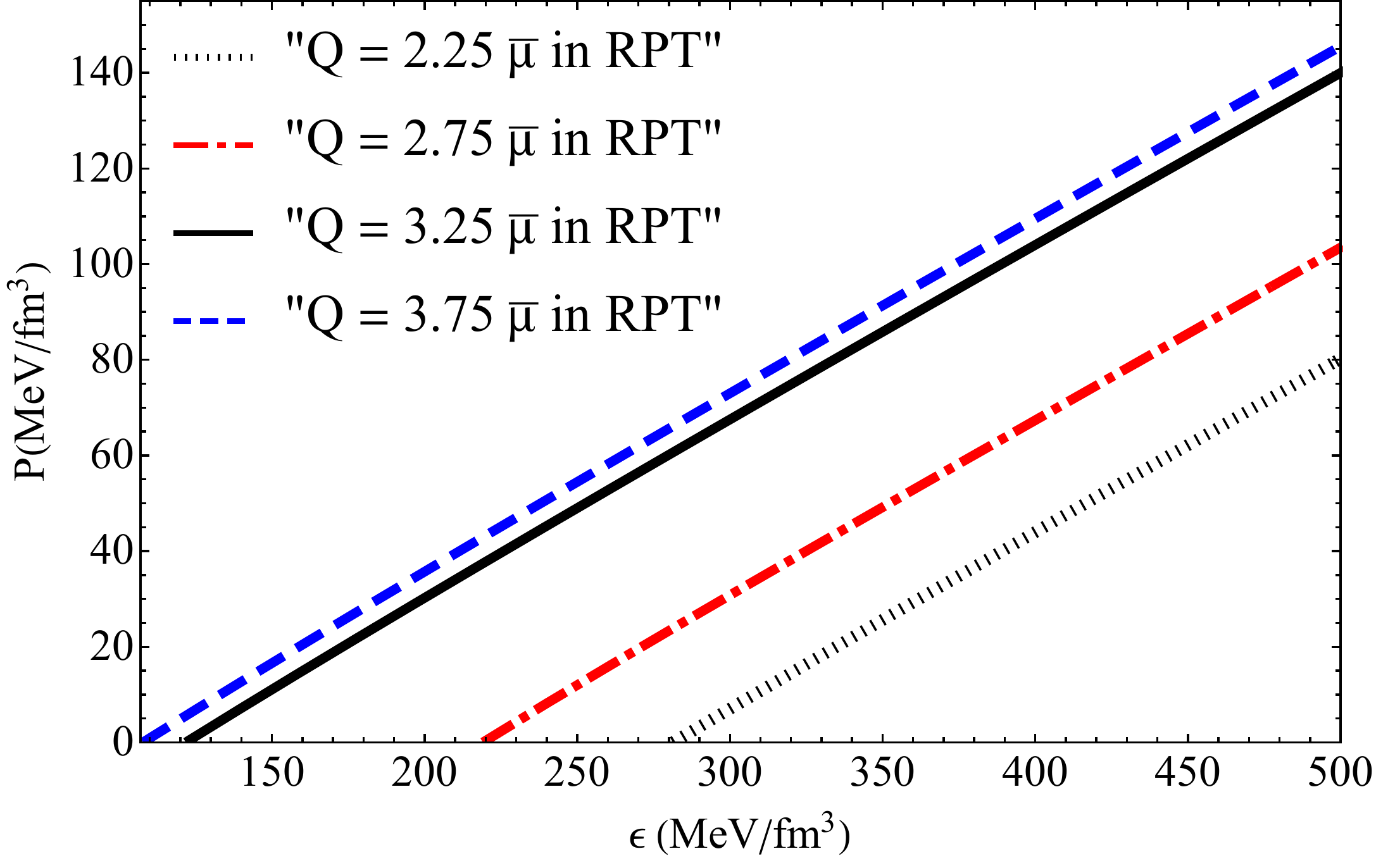}}
\caption{{\protect\small {Pressure versus energy density (EoS) for different
choices of renormalization scale in \textbf{RPT}}.}}
\label{EoS-in-RPT}
\end{figure}

\begin{figure}[h]
\label{speedRQCD} \center{\includegraphics[width=8.5cm]
		{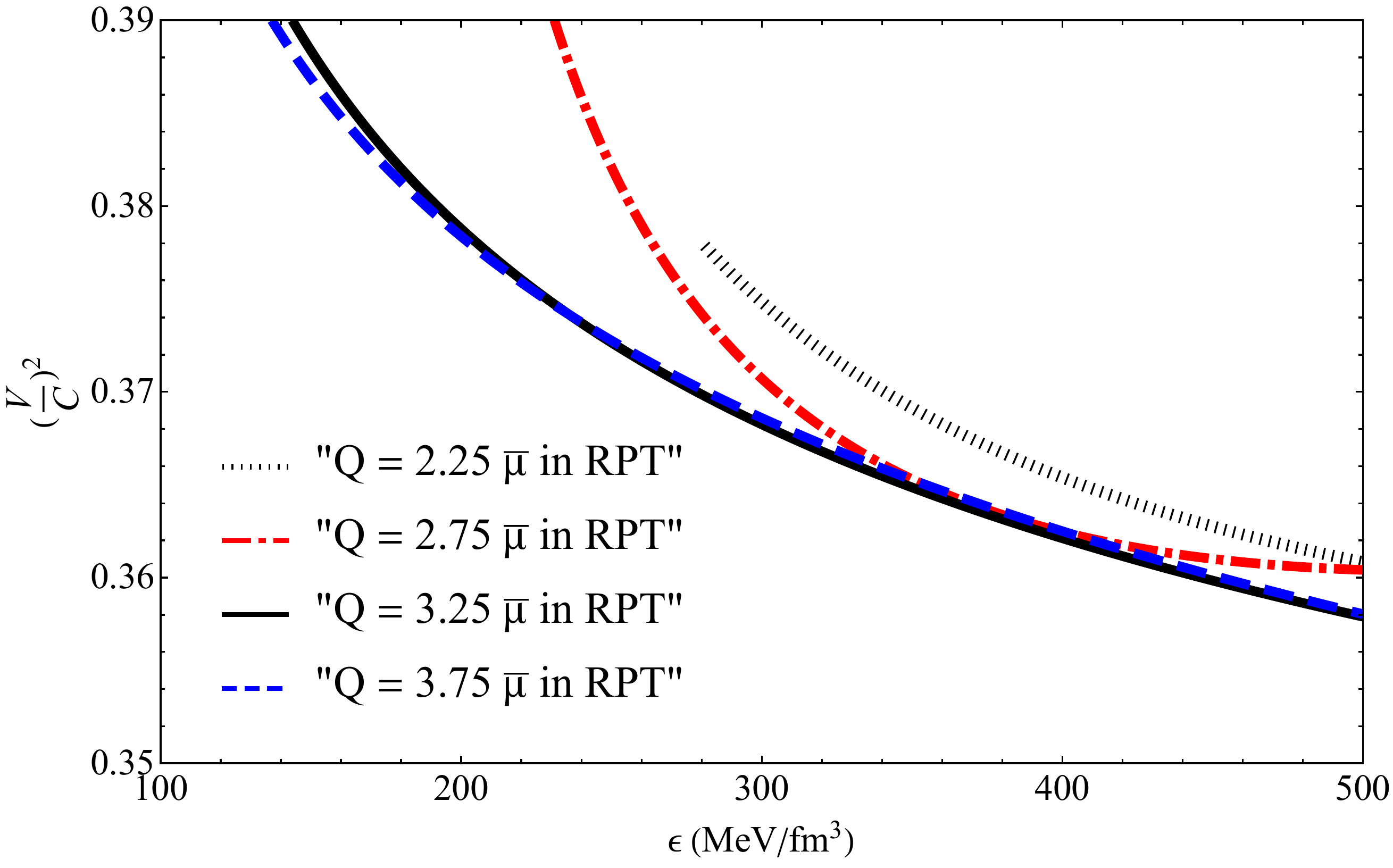}}
\caption{{\protect\small {\ Squared speed of sound versus energy density for
different choices of renormalization scale in \textbf{RPT}.}}}
\label{speed of sound in RPT}
\end{figure}

Using Eqs. (\ref{density}), (\ref{free}) and (\ref{omegatotal2}), the quark number
densities as functions of quark chemical potentials are obtained. Then, we
use Eqs. (\ref{24}), (\ref{25}), (\ref{26}) and (\ref{nB}) to get the
allowed values of chemical potentials for the quarks. The values of $%
\mathfrak{B}$ in Eq. (\ref{p}) are chosen in such a way to get zero total
pressure at the surface of the star where the baryon density is minimum. If
we neglect interaction between quarks, $\mathfrak{B}$ has the role of MIT
bag constant ranging from $41.58 \ MeV/fm^{3}$ to $319.31\ MeV/fm^{3}$ \cite%
{aziz2019constraining}. The EoS of SQM for different choices of
renormalization scale is shown in Fig. \ref{EoS-in-RPT}. The minimum value for the renormalization scale starts at a point from which the perturbative calculation is valid. Furthermore, for renormalization scales more than $Q=3.25\overline{\mu }$, the results do not change significantly. From this figure, it is understood that the EoS becomes stiffer by increasing the
renormalization scale. To establish causality condition, the ratio of the
speed of sound to the speed of light in vacuum $\left( \sqrt{dP/d\epsilon }%
\right) $ must be lower than unity. Fig. \ref{speed of sound in RPT}
shows that our results satisfy the causality condition for different choices
of renormalization scale.

\begin{figure}[h]
\center{\includegraphics[width=8.5cm]
		{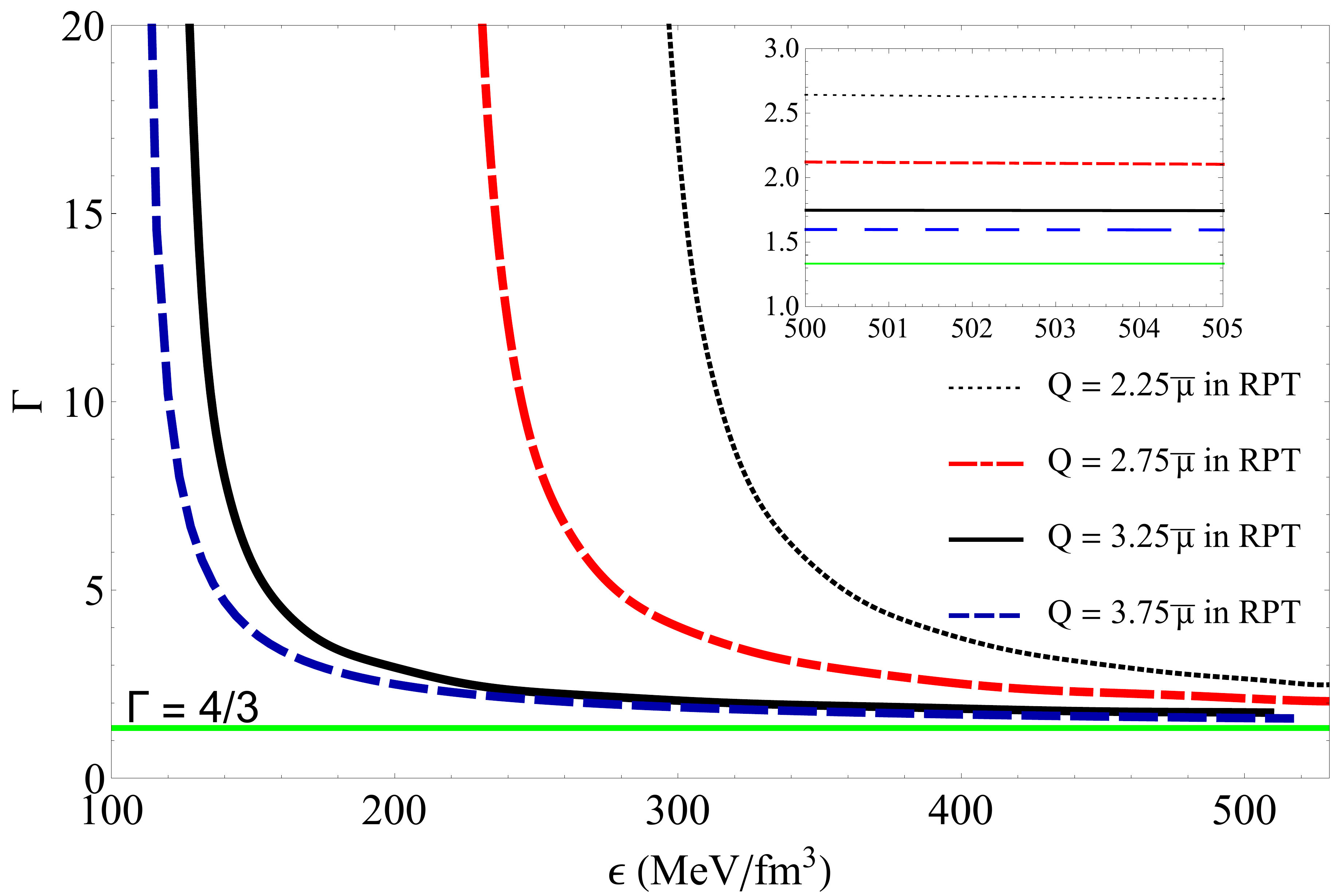}}
\caption{{\protect\small {Adiabatic index versus the ratio of baryon density
to nuclear saturation density for different choices of renormalization scale
in \textbf{RPT}.}}}
\label{adia RPT}
\end{figure}

It is known that for dynamical stability the adiabatic index $\Gamma
=dP/d\epsilon \dfrac{(P+\epsilon )}{P}$ must be higher than $4/3$ \cite%
{bardeen1966catalogue,knutsen1988stability,mak2013isotropic,panah2019white,panah2019contraction}%
. The adiabatic index has
been presented as a function of energy density in Fig. \ref{adia RPT}. As one can see from this figure, the condition \textcolor{red}{ $\Gamma > 4/3$} is satisfied  for all renormalization scales.

\subsection{Results in \textbf{BPT}}

Now we perform the calculations for \textbf{BPT} the same as we did for \textbf{RPT}.
Here, $M_{B}$ in Eq. (\ref{5}) is equal to $\sqrt{2\pi \sigma }$. The parameter $%
\sigma $ is the string tension which is equal to $0.18\ GeV^{2}$ \cite{deur2016qcd}.
Assuming the same constraints presented in section \ref{sec 5}, we obtain the EoS of
the SQM for different renormalization scales. For the values $Q>2.75\overline{\mu }$, the results for the structural properties of the star do not change significantly.
\begin{figure}[h]
\center{\includegraphics[width=8.5cm]
		{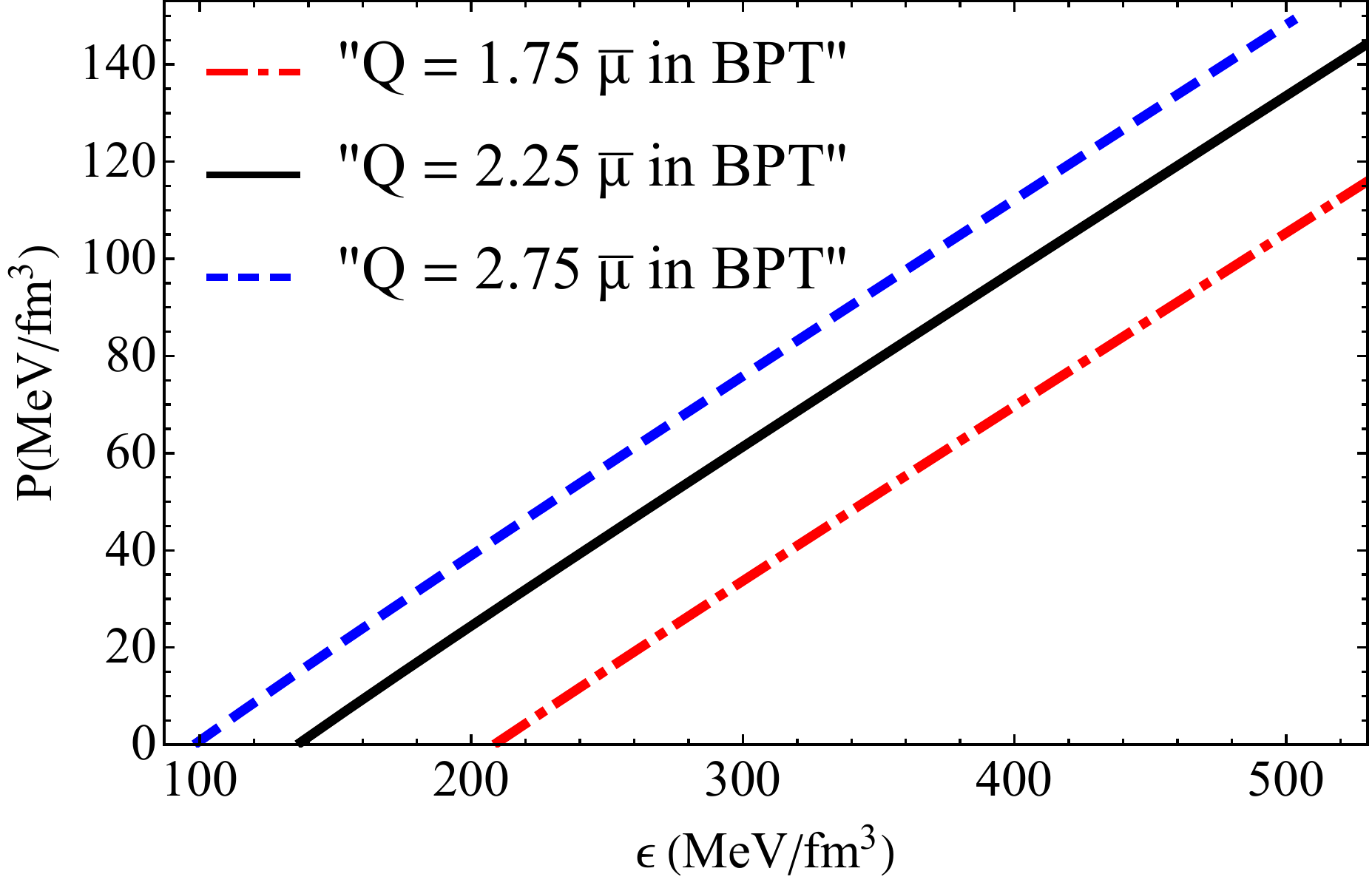}}
\caption{{\protect\small {Pressure versus energy density (EoS) for different
choices of renormalization scale in \textbf{BPT}}.}}
\label{EoS in RPT}
\end{figure}

\begin{figure}[h]
\center{\includegraphics[width=8.5cm]
		{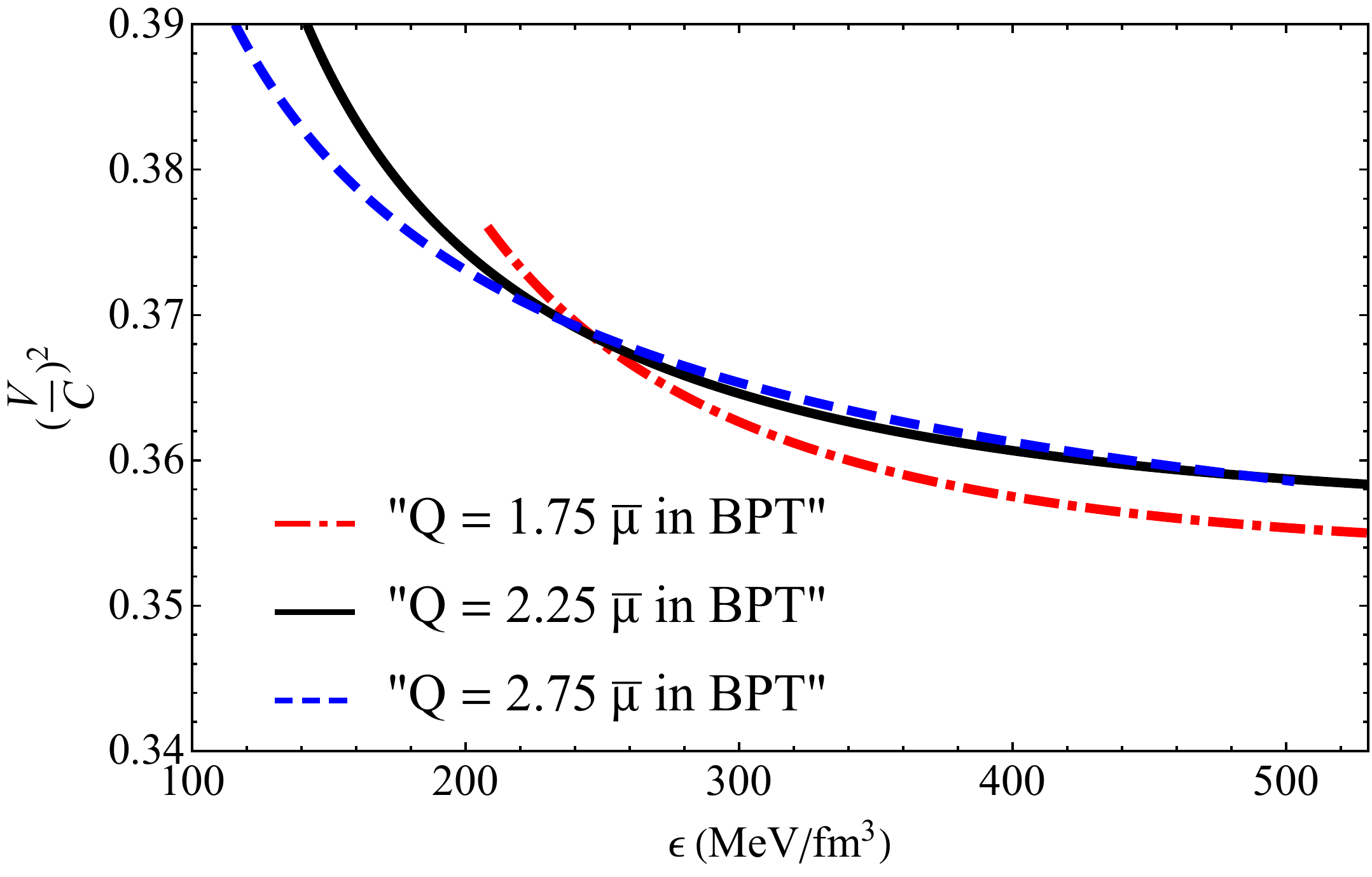}}
\caption{{\protect\small {Squared speed of sound versus energy density for
different choices of renormalization scale in \textbf{BPT}.} }}
\label{speed of sound in BPT}
\end{figure}

\begin{figure}[h]
\center{\includegraphics[width=8.5cm]
		{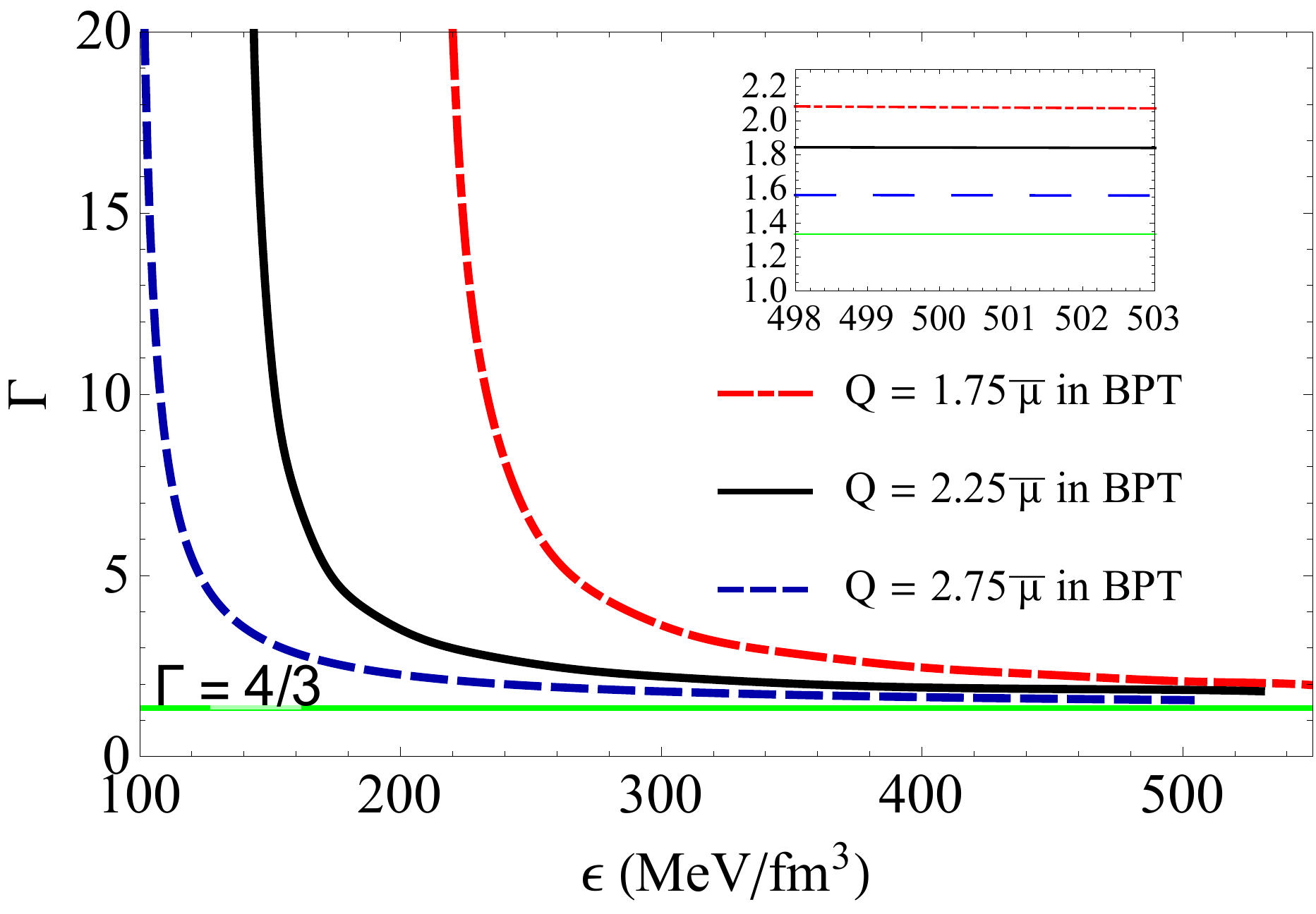}}
\caption{{\protect\small {\ Adiabatic index versus the ratio of baryon
density to nuclear saturation density for different choices of
renormalization scale in \textbf{BPT}.} }}
\label{adiaBPT}
\end{figure}

The diagrams of the speed of sound and adiabatic index in \textbf{BPT}
are shown in Figs. \ref{speed of sound in BPT} and \ref{adiaBPT}. These diagrams
indicate that the constraints related to the speed of sound and adiabatic indices are
satisfied well by our obtained EoSs.

\section{Magnetized strange quark star structure}

After computing the EoS, the structure of a hydrostatic and non-rotating
strange stars can be obtained by general relativistic equations of
hydrostatics. Here, the relation between mass and radius of the star is
found by solving TOV equations \cite{oppenheimer1939massive},
\begin{eqnarray}
\ dM(r) &=&4\pi r^{2}\epsilon (r)dr, \\
&&  \notag \\
\frac{dP(r)}{dr} &=&\dfrac{G\left[ P(r)+\epsilon (r)\right] \left[ M(r)+4\pi
r^{3}P(r)\right] }{r\left( 2GM(r)-r\right) },
\end{eqnarray}%
where $G$ and $r$ are the Newton gravitational constant and radial
coordinate of the star, respectively.
For different central pressure values, different gravitational masses
and radii are obtained. In the following, the results for different values of the renormalization
scale in  \textbf{RPT} and \textbf{BPT} approaches are discussed.

\begin{figure}[h]
\center{\includegraphics[width=8.5cm]
		{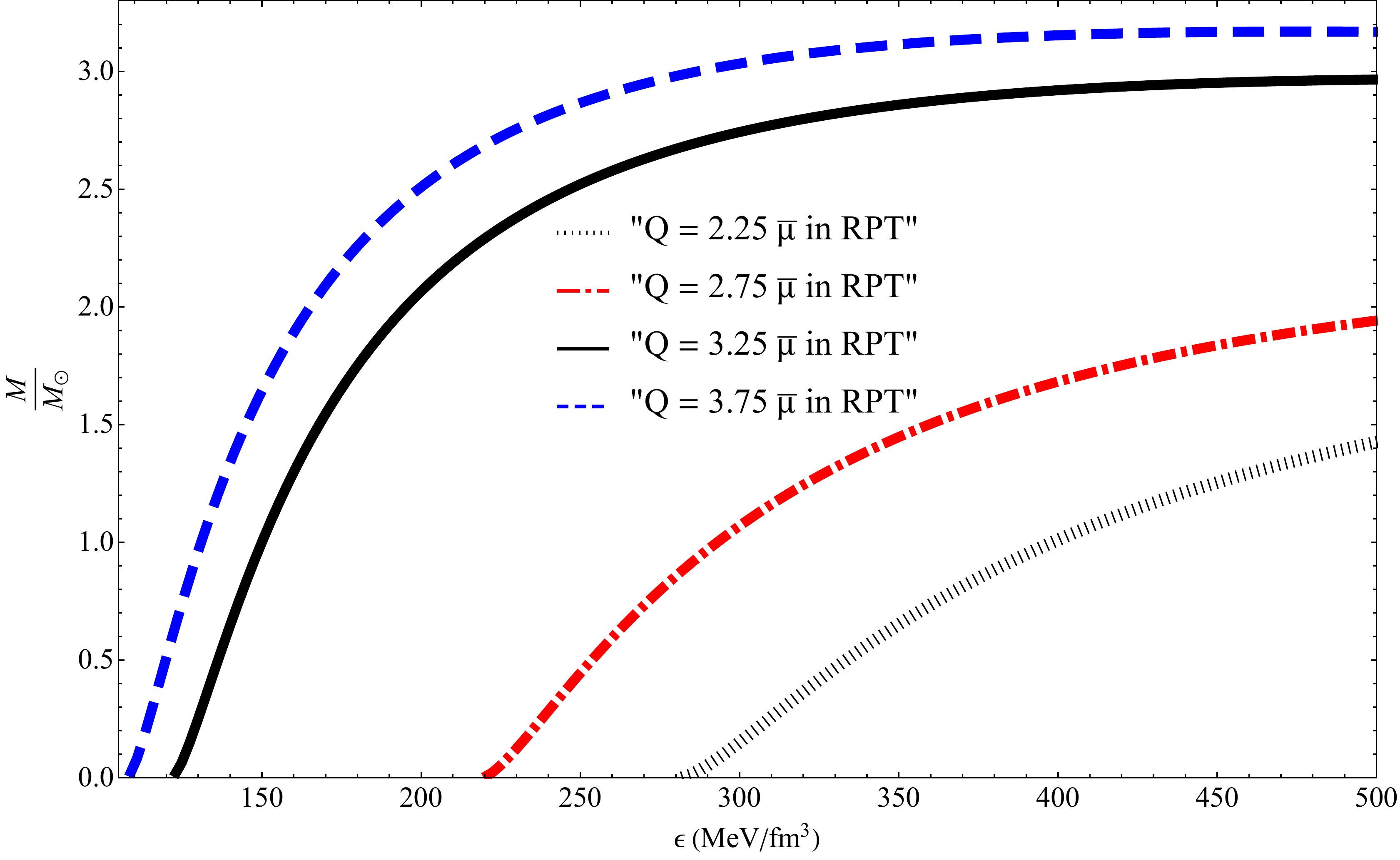}}
\caption{{\protect\small {\ Mass versus energy density for different choices
of renormalization scale in \textbf{RPT}.}}}
\label{mass in RPT}
\end{figure}

\begin{figure}[h]
\center{\includegraphics[width=8.5cm]
		{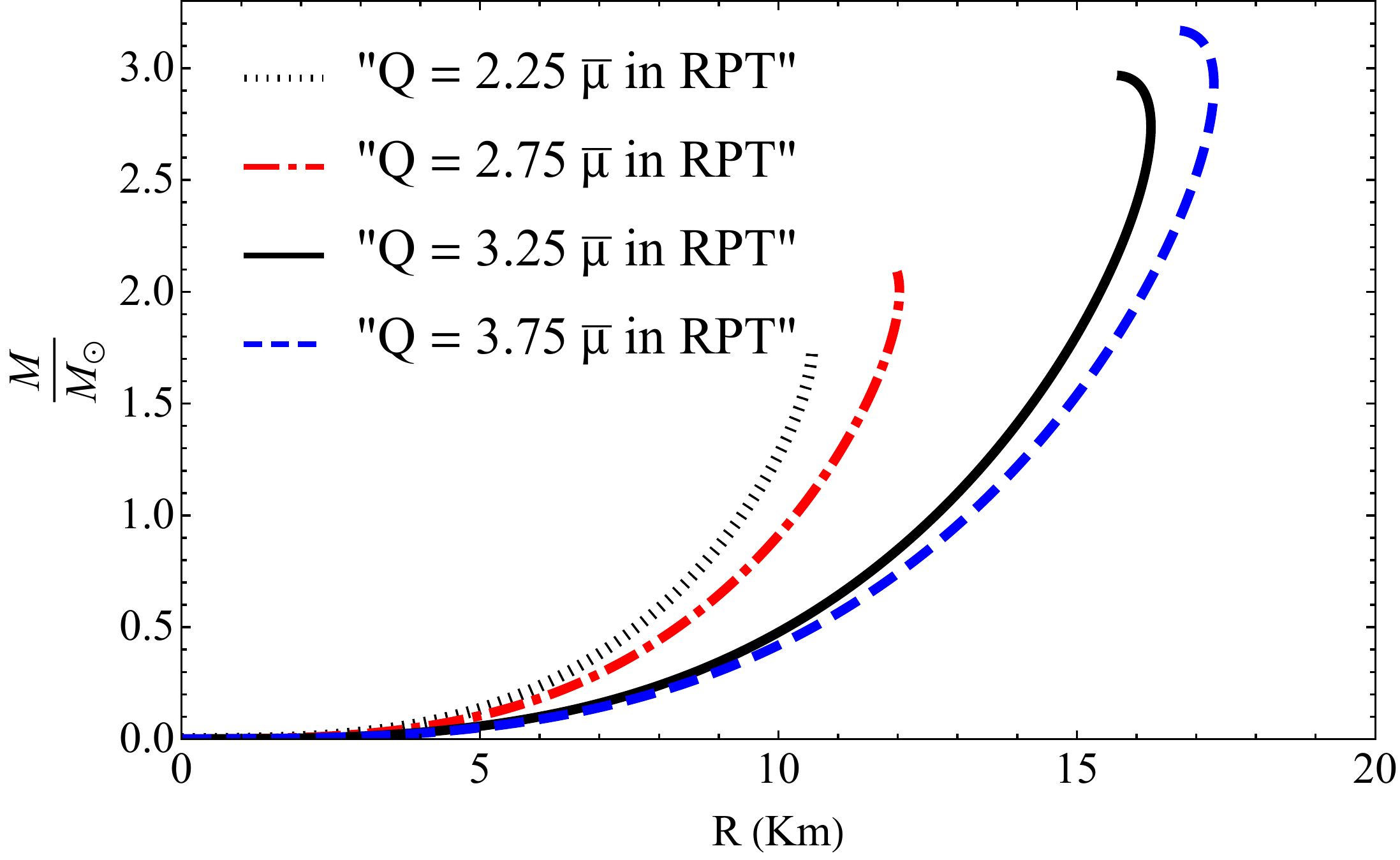}}
\caption{{\protect\small {\ Mass-radius diagram for different choices of
renormalization scale in \textbf{RPT}. }}}
\label{M&R in RPT}
\end{figure}

Our results for the structural properties of SQS in \textbf{RPT} have been
presented in Figs. \ref{mass in RPT} and \ref{M&R in RPT}. The limiting
behavior of mass in Fig. \ref{mass in RPT} shows the maximum gravitational
mass of SQS. From this figure, we see that the maximum gravitational mass
increases by increasing the renormalization scale, $Q$, which corresponds to
stiffer EoSs (Fig. \ref{EoS in RPT}). This behavior indicates that
stiffer EoS leads to higher gravitational mass for SQS in our model. Our
results for maximum gravitational mass and corresponding radius in \textbf{%
	RPT} are given in Table \ref{table-mass in RPT}. Previously
the maximum gravitational mass of the SQS by perturbative calculation in the absence of magnetic field was obtained in Refs. \cite%
{fraga2005role,kurkela2010cold}. In Ref. \cite{kurkela2010cold}, by using regular perturbation theory up to second order of coupling constant,  the maximum mass of a pure strange quark star was obtained to be $2.75M_{\odot}$. From Table. \ref%
{table-mass in RPT}, we see that in \textbf{RPT} model, the maximum mass of SQS can reach the value $M\simeq 3.17M_{\odot}$ which is considerably higher than that of in the absence of magnetic field calculated in Refs. \cite{kurkela2010cold,fraga2005role}.
\begin{table}[h]
\caption{{\protect\small {maximum gravitational mass and the corresponding radius of SQS for different renormalization scales in \textbf{RPT}.}}}
\label{table-mass in RPT}\centering
\begin{tabular}{||c|c|c|c|c|c|c||}
\hline
$Q$/$\overline{\mu}$ & $\dfrac{M}{M_\odot}$ & $R(km)$ & $R_{Sch} (km)$ & $z$
& $\dfrac{M_{BB}}{{M}_\odot}$ & $\sigma(10^{-1})$ \\ \hline
2.25 & 1.72 & 10.56 & 5.07 & 0.38 & 3.18 & 4.80 \\ \hline
2.75 & 2.09 & 11.98 & 6.16 & 0.43 & 3.61 & 5.14 \\ \hline
3.25 & 2.97 & 15.67 & 8.76 & 0.50 & 4.72 & 5.59 \\ \hline
3.75 & 3.17 & 16.53 & 9.34 & 0.52 & 4.98 & 5.65 \\ \hline
\end{tabular}%
\end{table}

\begin{figure}[h]
\center{\includegraphics[width=8.5cm]
		{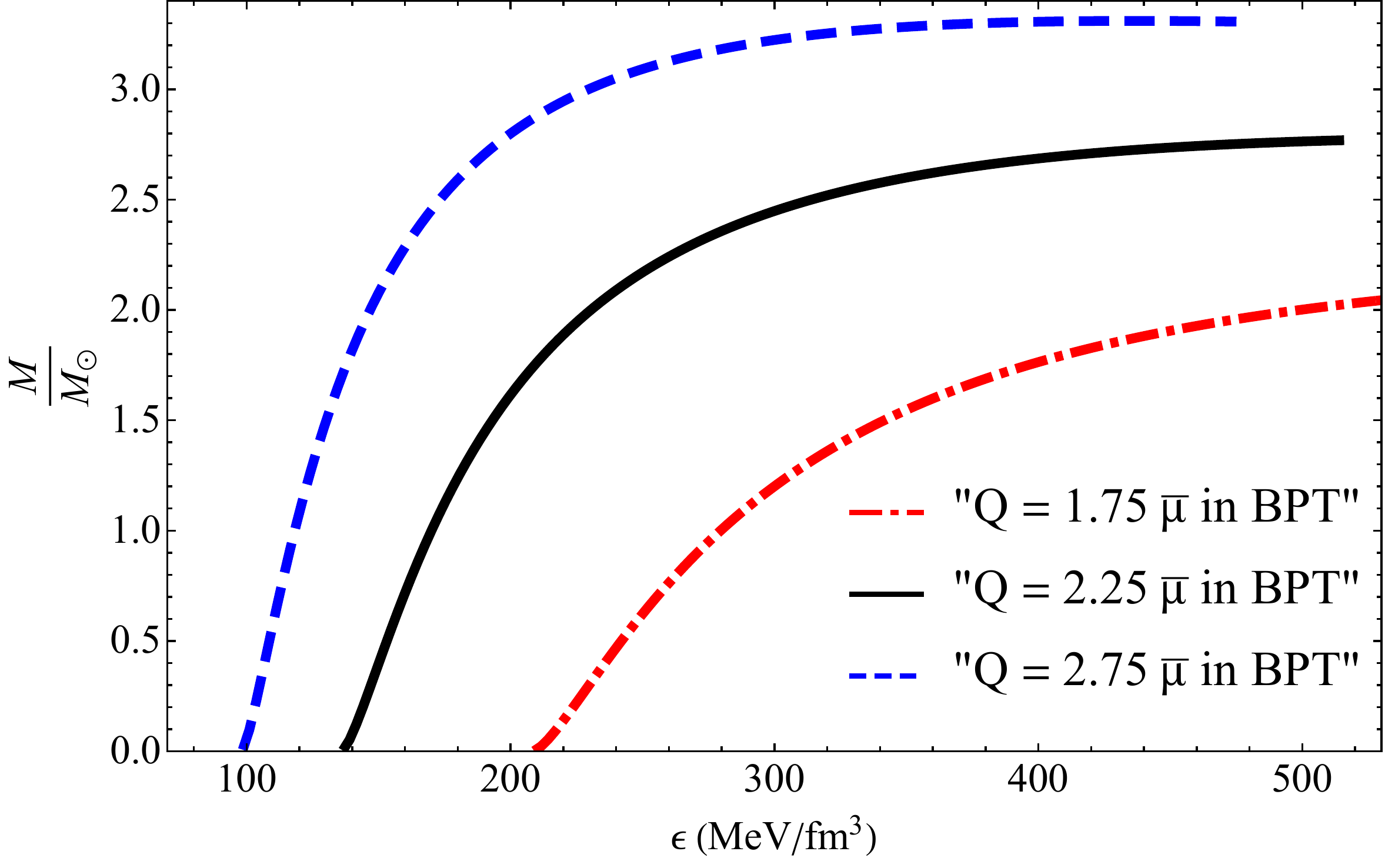}}
\caption{{\protect\small {\ Mass versus energy density for different choices
of renormalization scale in \textbf{BPT}.}}}
\label{mass in BPT}
\end{figure}

\begin{figure}[tbp]
\center{\includegraphics[width=8.5cm]
		{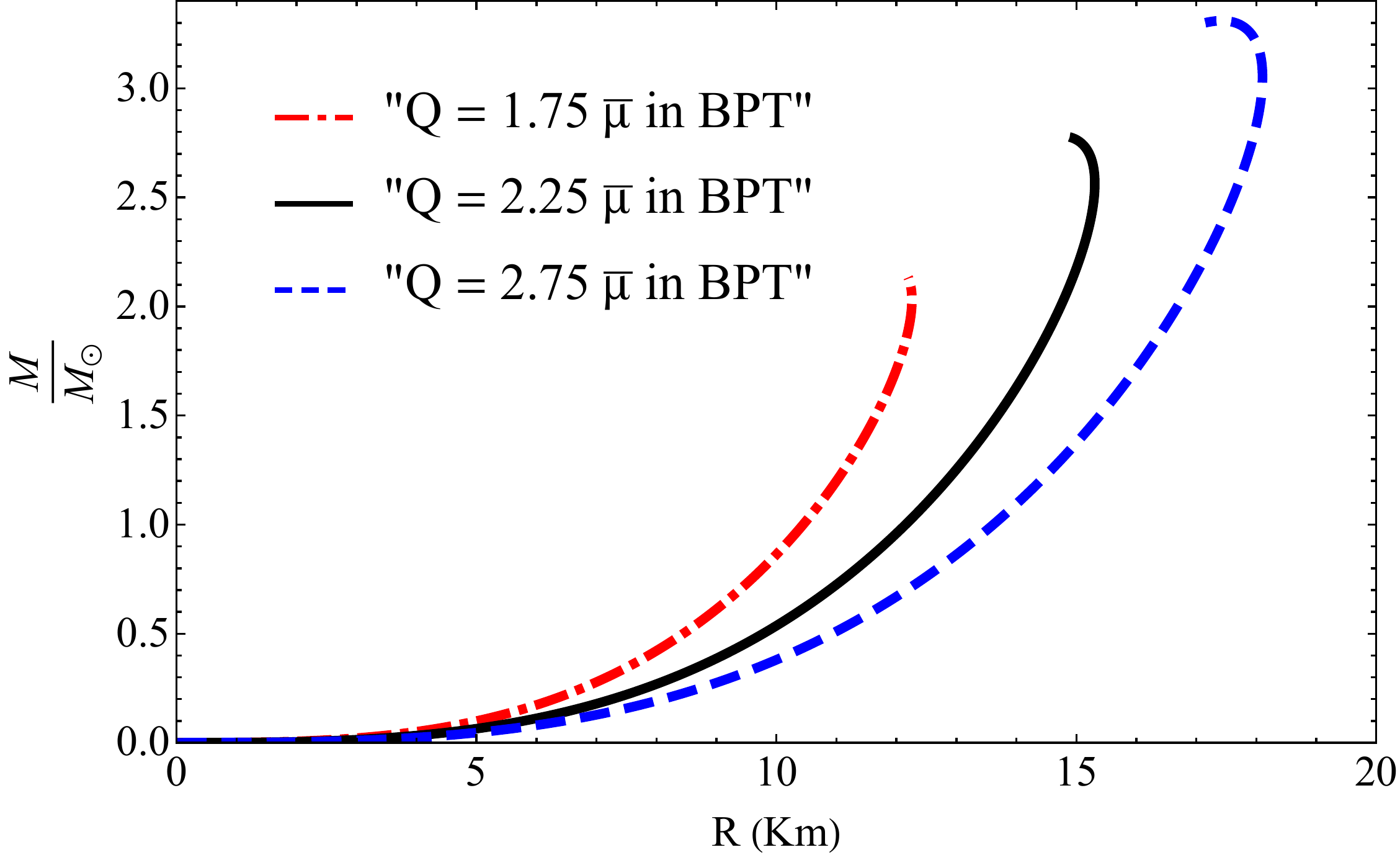}}
\caption{{\protect\small Mass-radius diagram for different choices of
renormalization scale in \textbf{BPT}.}}
\label{M&R in BPT}
\end{figure}

Now we discuss the maximum gravitational mass and the corresponding radius of the SQS in \textbf{%
BPT}.  Figs. \ref{mass in BPT} and \ref{M&R in BPT} show that by increasing $Q$, the maximum mass increases. Our
results for the maximum gravitational mass and the corresponding radius in
\textbf{BPT} are given in Table \ref{M&R BPT}. By comparing our results in
two approaches from Tables \ref{table-mass in RPT} and \ref{M&R BPT}, we can
conclude that the gravitational mass of strange quark star
in \textbf{BPT} is higher than that of \textbf{RPT}. It seems that it is originated from the behavior of the coupling constant. The coupling constant in \textbf{BPT} is smaller and runs slower than that in \textbf{RPT}.
\begin{table}[h]
\caption{{\protect\small {The values of the mass and the corresponding radius of the star in
different renormalization scales in \textbf{BPT}}}}
\label{M&R  BPT}\centering
\begin{tabular}{||c|c|c|c|c|c|c||}
\hline
$Q$/$\overline{\mu}$ & $\dfrac{M}{M_\odot}$ & $R(Km)$ & $R_{Sch} (km)$ & $z$
& $\dfrac{M_{BB}}{{M}_\odot}$ & $\sigma(10^{-1})$ \\ \hline
1.75 & 2.11 & 12.22 & 6.22 & 0.43 & 3.68 & 5.09 \\ \hline
2.25 & 2.77 & 14.96 & 8.17 & 0.48 & 4.51 & 5.46 \\ \hline
2.75 & 3.31 & 17.26 & 9.76 & 0.52 & 5.20 & 5.65 \\ \hline
\end{tabular}%
\end{table}

\subsection{Compactness}

Compactness is a quantity that gives us information about the strength of gravity of compact objects. The compactness of a spherical object is usually defined in the  following form
\begin{equation}
\sigma =\frac{R_{Sch}}{R},
\end{equation}%
where $R_{Sch}$ and $R$\ are the Schwarzschild radius and radius of compact
object, respectively. Our results for the compactness are presented in
Tables. \ref{table-mass in RPT} and \ref{M&R BPT}, which indicate that by
increasing renormalization scale ($Q$), the compactness from the perspective
of a distant observer (or an observer outside the compact object) increases.

\subsection{Gravitational Redshift}

Another quantity that gives us information on the strength of gravity is
related to the gravitational redshift. The gravitational redshift is given
by
\begin{equation}
z=\frac{1}{\sqrt{1-\frac{2GM}{c^{2}R}}}-1,
\end{equation}
where $M$ and $R$ are related to the compact star's mass and radius, respectively.
The results related to the gravitational redshift are given in Tables. \ref%
{table-mass in RPT} and \ref{M&R BPT}. These results indicate that the gravitational redshift increases by
increasing the renormalization scale. But
this quantity is less than $1$ for magnetized SQSs.

\subsection{Buchdahl-Bondi Bound}

Here, we want to study the upper mass limit of a static spherical quark star
with uniform density in GR, the so-called Buchdahl theorem \cite%
{buchdahl1959general}. The GR compactness limit is given by \cite%
{buchdahl1959general}
\begin{equation}
M\leq M_{BB}=\frac{4c^{2}R}{9G},  \label{BB}
\end{equation}%
in which the upper mass limit is $M_{\max }=\frac{4c^{2}R}{9G}$. Our
numerical results in Tables. \ref{table-mass in RPT} and \ref{M&R BPT},
confirm that the obtained masses of magnetized SQSs in GR respects the
equation (\ref{BB}).

Our calculations from the redshift ($z$)\ and Buchdahl-Bondi bound ($M_{BB}$%
) of magnetized SQSs confirm that these compact objects cannot be black
holes, because the values of redshift for these compact objects are finite ($%
z<1$),\ and also the maximum mass of these compact objects are less than
Buchdahl-Bondi bound ($M_{\max }<M_{BB}$).

\section{The effect of the anisotropy on the structure of the magnetized SQS}

\label{anisotropic structure}

In this section, we consider the magnetic fields with the values $10^{18}\ G$%
, $2\times 10^{18}\ G$ and $3\times 10^{18}\ G$ and calculate the maximum
mass of the SQS by considering longitudinal pressure, $P_{\lVert }$ and the
transverse pressure, $P_{\bot }$, in TOV equations separately. The maximum gravitational mass and the corresponding radius evaluated by $
P_{\lVert }$, are denoted by $M_{\lVert }$ and $R_{\lVert }$ and the ones obtained by $%
P_{\bot }$, are denoted by $M_{\bot }$ and $R_{\bot }$, respectively. Fig. \ref%
{aniso} and Table. \ref{aniso table} show our results in \textbf{RPT} by
assuming the renormalization scale $Q=3.75\overline{\mu }$. As we can see
from the results, there is no distinguishable difference between $M_{\lVert
} $ and $M_{\bot }$ for the magnetic field $10^{18}\ G$ and $2\times
10^{18}\ G $. But for $B=3\times 10^{18}\ G$ this difference is obvious and
the isotropic structure is not a good approximation. We have defined the $%
\delta $ parameter as the measure of the anisotropy of the star which is
equal to $\dfrac{M_{\lVert }-M_{\bot }}{\left( M_{\lVert }+M_{\bot }\right)
/2}$ \cite{chu2018quark}. The values of the $\delta $ parameter for
different magnetic fields have been presented in Table. \ref{aniso table}.
As we can see from this table, this parameter is negligible for the magnetic
fields less than about $2\times 10^{18}\ G$. The threshold magnetic field from
which an anisotropic approach begins to be  significant lies in the interval $2\times
10^{18}G<B<3\times 10^{18}G$. Such behavior can also be seen in  Ref.
\cite{chu2018quark} within the extended confined isospin-density-dependent
mass model. Therefore, it seems that considering the isotropic structure for a SQS under
the magnetic fields  $B<2\times 10^{18}G$ is reasonable in our models.
\begin{figure}[h]
\center{\includegraphics[width=8.5cm]
		{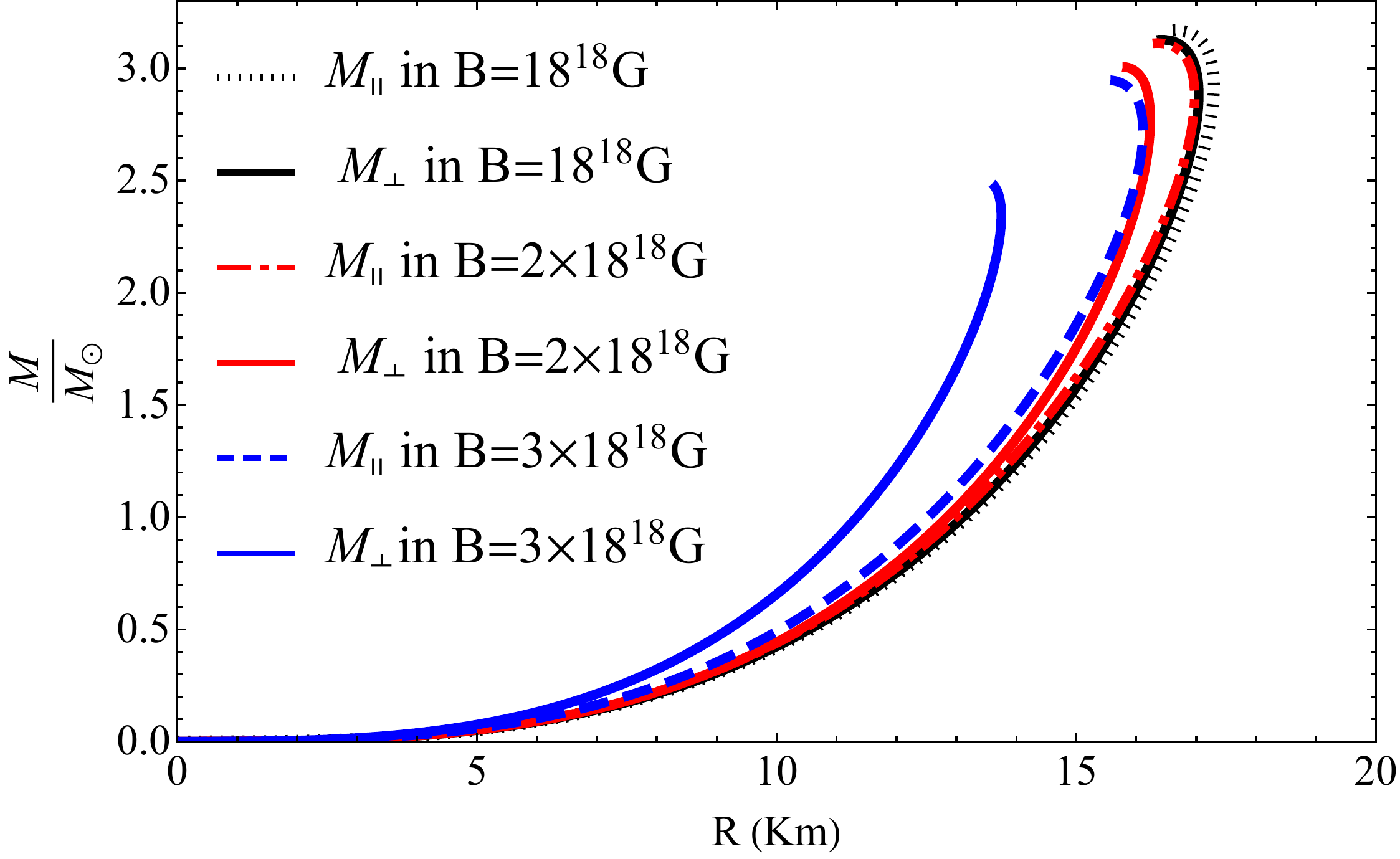}}
\caption{{\protect\small {\ Mass-radius diagram for different magnetic
fields in \textbf{RPT} and $Q=3.75\overline{\protect\mu }$. $M_{\lVert }$
and $M_{\bot }$ are the maximum mass calculated by the pressures $P_{\lVert
} $ and $P_{\bot }$, respectively.}}}
\label{aniso}
\end{figure}

\begin{table}[h]
\caption{{\protect\small {The values of the maximum mass and the corresponding radius of
the SQS for different magnetic fields in \textbf{RPT} and $Q=3.75\overline{%
\protect\mu }$. $M_{\lVert }$ and $M_{\bot }$ are the maximum mass
calculated by the pressures $P_{\lVert }$ and $P_{\bot }$, respectively. $%
R_{\lVert }$ and $R_{\bot }$ are their corresponding radii.}}}
\label{aniso table}\centering
\begin{tabular}{||c|c|c|c|c|c||}
\hline
$B(G)$ & $\dfrac{M_{\lVert }}{M_\odot}$ & $R_{\lVert }(Km)$ & $\dfrac{%
M_{\bot }}{M_\odot}$ & $R_{\bot }(Km)$ & $\delta(\%)$ \\ \hline
$10^{18}$ & 3.17 & 16.53 & 3.13 & 16.35 & 1.27 \\ \hline
$2\times10^{18}$ & 3.11 & 16.16 & 3.01 & 15.67 & 3.27 \\ \hline
$3\times10^{18}$ & 2.95 & 15.57 & 2.49 & 13.40 & 17 \\ \hline
\end{tabular}%
\end{table}

\section{Conclusions and Outlook}

In this paper, we first introduced QCD Feynman rules in the presence of a strong magnetic field. Then we performed perturbative calculations of the thermodynamic potential of a strange quark matter (SQM) up to \emph{O}($\alpha _{s}$) in the presence of magnetic field $B=10^{18}\ G$. We used two approaches in our perturbative calculations: i) QCD without IR freezing effect in regular perturbation theory (\textbf{RPT}) and ii) QCD with IR freezing effect in background
perturbation theory (\textbf{BPT}). The thermodynamic potential of SQM was computed by separating it into non-perturbative and perturbative parts.
The non-perturbative part contained the interaction of quarks with the magnetic field, and the perturbative part included QCD interaction between quarks.
%By using the Feynman rules in the presence of a magnetic field, the thermodynamic
The perturbative part contained a two-loop diagram added to the thermodynamic potential of a moving fermion in a uniform magnetic field as the non-perturbative part.
Using the obtained thermodynamic potential and also considering the stability conditions of SQM, thermodynamic properties
such as quark number densities, speed of sound, adiabatic index, and also the EoSs of
strange quark stars (SQSs) were obtained in \textbf{RPT} and \textbf{BPT} models. Then by employing the obtained EoSs, we calculated the structure properties of SQSs. Moreover, the redshift,
compactness, and Buchdahl-Bondi bound of the SQSs were calculated to show that these compact
objects cannot be black holes. Then, the anisotropic effect on the structure of SQS in our models under different magnetic fields was calculated. it was shown that the threshold magnetic field from which an anisotropic
approach begins to be significant lies in the interval $2\times
10^{18}G<B<3\times 10^{18}G$. The maximum mass of the SQS under the magnetic
field $B=10^{18}\ G$ was obtained $3.17M_{\odot }$ and $3.31M_{\odot }$ in
\textbf{RPT} and \textbf{BPT}, respectively, which is considerably higher
than those of Bag models and NJL theories.  \textcolor{red}{Furthermore, it is notable that some of our results for the maximum gravitational mass are outside the limits obtained by observational constraints for neutron stars. Therefore an observed violation of those limits might be a piece of evidence for  SQS. For example the compact object with the mass $2.5 - 2.67 M_{\odot }$ in GW190814 \cite{miao2021bayesian} or the remnant mass of GW190425 \cite{sedaghat2021compact} might be SQSs.}

\section*{Acknowledgements}
We wish to thank Shiraz University Research Council. S. M. Zebarjad thanks
Physics Department of \textbf{UCSD} for hospitality during his sabbatical.
B. Eslam Panah thanks the University of Mazandaran.
We are indebted to Yu.A. Simonov for his good comments and discussions, and
A. Manohar for reading the manuscript and his helpful comment.

\section{Appendix}

\appendix

\section{Calculation of the thermodynamic potential of the two-loop diagram}\label{appendixA}
By using the Feynman rules introduced in section \ref {sec 3} we have
\begin{align}
 \dfrac{\Omega _{2L}}{V}=&\dfrac{g^{2}d_{A}}{2}  \notag  \label{omega} \\
& \times\sum_{n,j=0}^{\infty }\int \dfrac{d^{4}p}{(2\pi )^{4}}\int \dfrac{%
	d^{4}k}{(2\pi )^{4}}\dfrac{(-1)^{j+n}e^{-(p_{\perp }^{2}+k_{\perp
		}^{2})/{\lvert e_{f}B\rvert }}}{(p-k)^{2}}  \notag
\notag \\
&\times  \dfrac{Tr\left[ \gamma ^{\mu }D_{j}(\lvert e_{f}B\rvert ,p)\gamma
	_{\mu }D_{n}(\lvert e_{f}B\rvert ,k)\right] }{\mathcal{A}_{1}\mathcal{A}_{2}}%
,
\end{align}%
where $\mathcal{A}_{1}=(p_0+i\mu)^{2}+(p_{3}^{2}+m_{f}^{2}+2j\lvert e_{f}B\rvert
) $ and $\mathcal{A}_{2}=(k_0+i\mu)^{2}+(k_{3}^{2}+m_{f}^{2}+2n\lvert
e_{f}B\rvert) $. Also, $d_{A}$ is $N_{c}^{2}-1$. The summation is over Landau levels, and $D_{j}(\lvert e_{f}B\rvert ,p)$ and $D_{n}(\lvert e_{f}B\rvert ,k)$ are functions in terms of generalized Laguerre polynomials and Dirac Gamma matrices in Euclidean space.
To obtain the contribution (\ref{omega}), we use a very useful method derived in Ref. \cite{ghicsoiu2017high} which is referred to as cutting rules. Based on this method, the evaluation of the Feynman integrals at zero temperature and finite chemical potential is reduced to the evaluation of three-dimensional phase space integrals over on-shell amplitudes at $T=\mu=0$. Indeed, the one point irreducible \emph{N}-loop \emph{n}-point Feynman diagram,  $F(P_k,\mu)$,  can be simplified as:
\begin{align}
&F(P_k,\mu)=\\ \notag &F_{0- cut}(P_k)+F_{1-cut}(P_k,\mu)+...+F_{N-cut}(P_k,\mu),
\end{align}
where $F_{0- cut}(P_k)$ is the original diagram computed at zero chemical potential, and the other terms are contributions resulting from the cutting procedure. Indeed, $F_{j-cut}(P_k,\mu)$ represents the sum of all diagrams in which the \emph{j} number of internal fermion propagators have been cut from the original graph. In the following, we explain the steps in the cutting procedure:

1. The cut propagators are removed from the
original

graph\\

2. The remaining $N-j$- loop $n+2j$- point amplitude

is computed by considering the point that the all

external momenta are real-valued. \\

3. The cut momenta $p_i$ is set on-shell by inserting

$({p_0})_i=iE_{i}$.\\

4. The remaining expression is integrated over
the cut

three-dimensional momenta $p_i$ with the weight

$\dfrac{-\theta(\mu-E_{i})}{2E_{i}}$, where for a quark moving in a uniform

magnetic field $B$ at the $j$th Landau level, $E_p$ is equal

to $\sqrt{p_{3}^{2}+m_{f}^{2}+2j\lvert e_{f}B\rvert}$.\\

Notably, $F_{0- cut}(P_k)$ has no dependence on chemical potential, and according to Eq. (\ref{density}), it has no contribution in quark number densities. Therefore, it is has been removed from calculations.
By following the cutting procedure and Eq. (\ref{omega}), we get
\begin{equation}\label{omegatotal}
\dfrac{\Omega _{2L}}{V}=F_{1-cut}+F_{2-cut},
\end{equation}
where $F_{1-cut}$ and $F_{2-cut}$ are given as
\begin{align}
F_{1-cut}& =-2\times \dfrac{g^{2}d_{A}}{2}\bigg\{\sum_{n,j=0}^{\infty }\int
\dfrac{\theta (\mu -E_{p})}{2E_{p}}\dfrac{d^{3}\mathbf{p}}{(2\pi )^{3}}
\notag \\
&  \notag \\
& \times \int \dfrac{d^{4}k}{(2\pi )^{4}}I_{1}\bigg\}_{p_{0}=iE_{p}}, \\
&  \notag \\
F_{2-cut}& =\dfrac{g^{2}d_{A}}{2}\bigg\{\sum_{n,j=0}^{\infty }\int \dfrac{%
	\theta (\mu -E_{p})}{2E_{p}}\dfrac{d^{3}\mathbf{p}}{(2\pi )^{3}}  \notag \\
&  \notag \\
& \times \int \dfrac{\theta (\mu -E_{k})}{2E_{k}}\dfrac{d^{3}\mathbf{k}}{%
	(2\pi )^{3}}I_{2}\bigg\}_{p_{0}=iE_{p},k_{0}=iE_{k}},
\end{align}%
where, $I_1$ and $I_2$ are defined as
\begin{align}
I_{1}& =(-1)^{j+n}e^{-(p_{\perp }^{2}+k_{\perp }^{2})/{\lvert e_{f}B\rvert }}
\notag \\
&  \notag \\
& \times \dfrac{Tr\left[ \gamma ^{\mu }D_{j}(\lvert e_{f}B\rvert ,p)\gamma
	_{\mu }D_{n}(\lvert e_{f}B\rvert ,k)\right] }{({k_{0}}^{2}+{E_{k}}%
	^{2})((p_{0}-k_{0})^{2}+(\mathbf{p}-\mathbf{k})^{2})}, \\
&  \notag \\
I_{2}& =(-1)^{j+n}e^{-(p_{\perp }^{2}+k_{\perp }^{2})/{\lvert e_{f}B\rvert }}
\notag \\
&  \notag \\
& \times \dfrac{Tr\left[ \gamma ^{\mu }D_{j}(\lvert e_{f}B\rvert ,p)\gamma
	_{\mu }D_{n}(\lvert e_{f}B\rvert ,k)\right] }{(p_{0}-k_{0})^{2}+(\mathbf{p}-%
	\mathbf{k})^{2}}.
\end{align}%
Due to the Landau quantization of transverse
momentums at the $j$th
Landau level, we set  $\sqrt{2j\lvert e_{f}B\rvert }$ and $\sqrt{%
	2(j+1)\lvert e_{f}B\rvert }$ for the lower and upper limits of the
transverse momentum integrals, respectively. To calculate Eq. (\ref{omegatotal}),
we need to sum over all the terms corresponding to $n$th and $j$th Landau levels. Indeed for each pair $(n, j)$  which appear as Landau levels in propagators of quarks, there is an integral which should be computed. It should be noted that by multiplying a factor $-1/2$ in Eq. (\ref{omegatotal}), originated from perturbative expansion \cite{kapusta2006finite}, we obtain
Eq. (\ref{omegatotal2}).
\begin{align}\label{omegatotal2}
& \dfrac{\Omega _{2L}}{V}=\dfrac{g^{2}d_{A}}{2}\sum_{n,j=0}^{\infty }\bigg\{\int \dfrac{d^{4}p}{(2\pi )^{4}}\int \dfrac{%
	d^{4}k}{(2\pi )^{4}}I_1(-\theta(\mu-E_q))\notag  \\&-\dfrac{1}{2}\int \dfrac{d^{4}p}{(2\pi )^{4}}\int \dfrac{%
	d^{4}k}{(2\pi )^{4}}I_2(-\theta(\mu-E_q))(-\theta(\mu-E_p))\bigg\}
,
\end{align}

\section{Investigating the effect of the gap parameter in maximum gravitational mass of SQS}\label{gaps}
\begin{figure}[h]
	\center{\includegraphics[width=8.5cm]
		{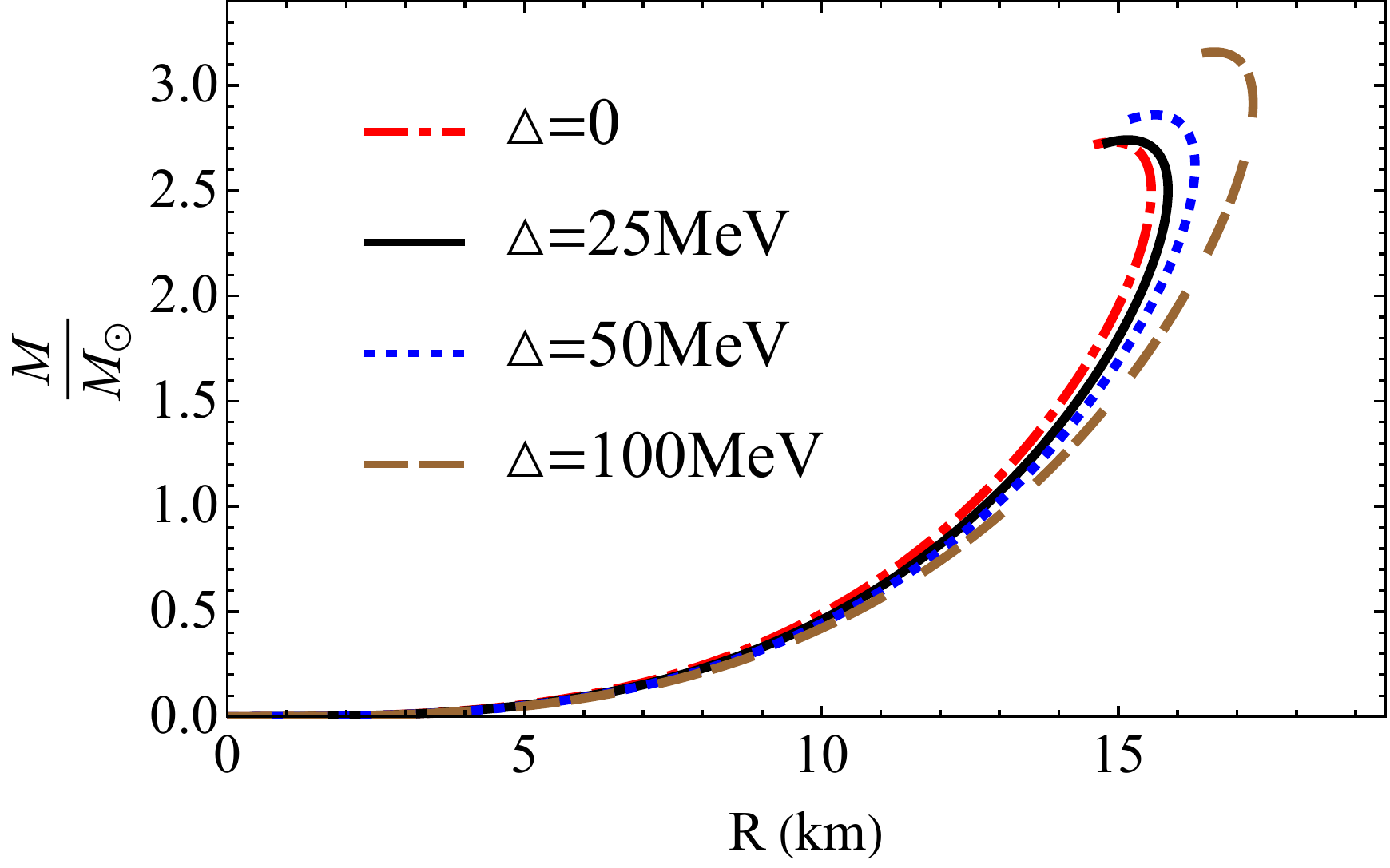}}
	\caption{{\protect\small{\ Mass-radius diagram for different values of $\Delta$ parameter in RPT ($Q=3.75\overline{\mu}$) model.}}}
	\label{gapRPT}
\end{figure}

\begin{figure}[h]
	\center{\includegraphics[width=8.5cm]
		{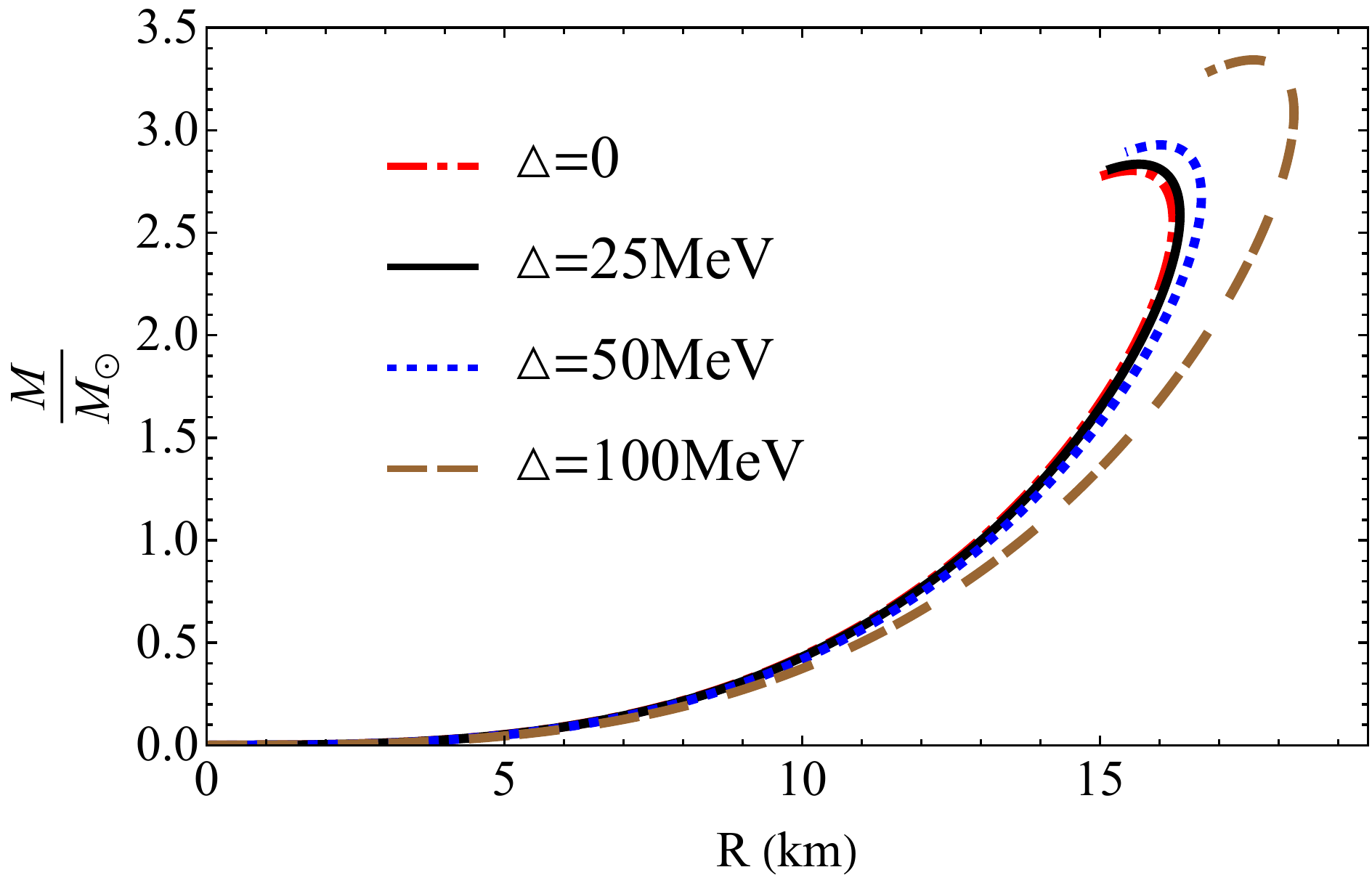}}
	\caption{{\protect\small {\ Mass-radius diagram for different values of $\Delta$ parameter in BPT ($Q=2.75\overline{\mu}$) model.}}}
	\label{gapBPT}
\end{figure}

By using different values of $\Delta$ in Eq. (\ref{p}), we have plotted mass-radius diagrams for SQS under the magnetic field $B=10^{18}G$. As the Table \ref{diffgap} and Figures \ref{gapRPT} and \ref{gapBPT} show, the maximum gravitational mass of SQS increases by increasing $\Delta$ for both \textbf{RPT} and \textbf{BPT} models. Furthermore, it is notable that our calculations show that the impact of $\Delta$ on the structure of SQS begins to be significant from  $\Delta > 25MeV$.
\begin{table}
	\caption{{\protect\small {The values of the maximum mass and the corresponding radius of SQS for different values of $\Delta$ in \textbf{RPT}($Q=3.75\overline{\mu}$) and \textbf{BPT}($Q=2.75\overline{\mu}$) models.}}}
	\label{diffgap}\centering
	\begin{tabular}{||c|cc|cc||}
		\hline
		$$ & RPT & RPT & BPT & BPT \\
		\hline
		$\Delta (MeV)$ & $\dfrac{M_{\lVert }}{M_\odot}$ & $R_{\lVert }(Km)$ & $\dfrac{%
			M_{\lVert }}{M_\odot}$ & $R_{\lVert }(Km)$ \\ \hline
	
		$0$ & 2.71 & 14.81 & 2.84 & 15.80 \\ \hline
		$25$ & 2.75 & 15.00 & 2.87 & 15.89 \\ \hline
		$50$ & 2.86 & 15.65 & 2.95 & 16.04 \\ \hline
		$100$ & 3.17 & 16.53 & 3.31 & 17.26\\ \hline
	\end{tabular}%
\end{table}
\pagebreak
\bibliographystyle{spphys}
\bibliography{refs}

\end{document}